\documentclass[10pt,a4paper]{article}
\pdfoutput=1
\usepackage{jheppub}
\usepackage{amsmath}
\usepackage{verbatim}
\usepackage{amssymb}
\usepackage{inputenc, array}
\usepackage{textcomp}
\usepackage{appendix}
\usepackage{amsfonts}
\usepackage{amsthm}
\usepackage{tikz}
\usetikzlibrary{decorations.markings}
\usepackage{mathtools}
\usepackage{bm}
\allowdisplaybreaks
\title{$w_{1+\infty}$ and Carrollian Holography}
\author{Amartya Saha}
\affiliation{Indian Institute of Technology Kanpur, Kanpur 208016, INDIA} 
\emailAdd{amartyas@iitk.ac.in}
\preprint{}
\abstract{In a $1+2$D Carrollian conformal field theory, the Ward identities of the two local fields $S^+_0$ and $S^+_1$, entirely built out of the Carrollian conformal stress-tensor, contain respectively up to the leading and the subleading positive helicity soft graviton theorems in the $1+3$D asymptotically flat space-time. This work investigates how the subsubleading soft graviton theorem can be encoded into the Ward identity of a Carrollian conformal field $S^+_2$. The operator product expansion (OPE) $S^+_2S^+_2$ is constructed using general Carrollian conformal symmetry principles and the OPE commutativity property, under the assumption that any time-independent, non-Identity field that is mutually local with $S^+_0,S^+_1,S^+_2$ has positive Carrollian scaling dimension. It is found that, for this OPE to be consistent, another local field $S^+_3$ must automatically exist in the theory. The presence of an infinite tower of local fields $S^+_{k\geq3}$ is then revealed iteratively as a consistency condition for the $S^+_2S^+_{k-1}$ OPE. The general $S^+_kS^+_l$ OPE is similarly obtained and the symmetry algebra manifest in this OPE is found to be the Kac-Moody algebra of the wedge sub-algebra of $w_{1+\infty}$. The Carrollian time-coordinate plays the central role in this purely holographic construction. The 2D Celestial conformally soft graviton primary $H^k(z,\bar{z})$ is realized to be contained in the Carrollian conformal primary $S_{1-k}^+(t,z,\bar{z})$. Finally, the existence of the infinite tower of fields $S^+_{k}$ is shown to be directly related to an infinity of positive helicity soft graviton theorems.}
\begin{document}
\maketitle
\flushbottom
\section{Introduction}
The thriving research program to understand the holographic principle \cite{tHooft:1993dmi,Susskind:1994vu} for the case of the $1+3$D asymptotically flat space-times (AFS) has been approached mainly via two\footnote{A third path to $1+3$D flat holography is to impose a `large AdS radius' limit on the appropriate results obtained in the framework of the AdS$_4$/CFT$_3$ holography. This approach, which is actually the oldest among the three, has seen recent resurgence through the works \cite{deGioia:2023cbd,Hijano:2020szl,deGioia:2022fcn,Bagchi:2023fbj}.} 
seemingly different avenues. The currently most well-developed one is known as the Celestial holography \cite{Pasterski:2021raf} where the dual Celestial CFT \cite{Pasterski:2016qvg} is thought to live on a Celestial sphere $S^2$ at the null-boundary of the AFS. The Celestial conformal fields depend on the two stereographic coordinates $(z,\bar{z})$ on the Celestial $S^2$ and are labeled by a Celestial conformal weight $\Delta_c$ which is a continuous parameter \cite{Pasterski:2017ylz}; thus, these fields effectively are functions of three variables. In the other framework known as the Carrollian holography, gaining attention in the recent years \cite{Bagchi:2016bcd,Donnay:2022aba,Bagchi:2022emh,Donnay:2022wvx,Saha:2023hsl,Salzer:2023jqv,Nguyen:2023vfz}, the dual Carrollian CFT (CarrCFT) resides on the three-dimensional null-infinity (with topology $\mathbb{R}\times S^2$) where the Carrollian conformal fields also depend on three variables: $(z,\bar{z})$ as well as the null-coordinate or the Carrollian time $t$. 

\medskip

The central idea of the Celestial holography \cite{Pasterski:2016qvg,Pasterski:2017ylz} is that the null-momentum space scattering amplitude of a mass-less scattering process in the $1+3$D bulk AFS can be re-expressed as a 2D Euclidean CFT correlation function, via a Mellin transformation that trades the energies $\{\omega\}$ of the external bulk scattering particles for the (continuous) Celestial conformal weights $\left\{\Delta_c\right\}$. Even before the explicit examples presented in \cite{Pasterski:2016qvg,Pasterski:2017ylz}, that the holographic dual of the (quantum) gravity-theory in the $1+3$D AFS might be a 2D Euclidean CFT was partly motivated by the realization that the Weinberg (leading) soft graviton theorem \cite{Weinberg:1965nx} and the Cachazo-Strominger subleading soft graviton theorem \cite{Cachazo:2014fwa} can be respectively recast into the 2D CFT U$(1)$ Kac-Moody \cite{Strominger:2013jfa} and the 2D CFT energy-momentum (EM) tensor Ward identities \cite{Kapec:2016jld}. These works were in turn inspired by linking up the following four observations: that the asymptotic symmetry group at the null-infinity of the $1+3$D AFS is the BMS$_4$ group \cite{Bondi:1962px,Sachs:1962wk} in the presence of the gravitational radiation, that the Weinberg soft graviton theorem \cite{Weinberg:1965nx} is completely equivalent to the BMS$_4$ super-translation Ward identity \cite{Strominger:2013jfa,He:2014laa}, that the Lorentz $\text{SL}(2,\mathbb{C})$ subgroup of the original BMS$_4$ group should be infinitely enhanced to include the 2D singular conformal transformations (super-rotations) on the Celestial $S^2$ \cite{Barnich:2009se,Barnich:2010ojg,Barnich:2011mi} and finally, that the Cachazo-Strominger subleading soft graviton theorem \cite{Cachazo:2014fwa} implies a Virasoro symmetry of the $1+3$D AFS gravitational $S$-matrix \cite{Kapec:2014opa}.

\medskip

On the other hand, the starting point of the Carrollian holography is the observation \cite{Duval:2014uva,Duval:2014lpa} that the original BMS$_4$ group \cite{Bondi:1962px,Sachs:1962wk} is isomorphic to the $1+2$D conformal Carroll group (at level 2) on a Carrollian manifold with topology $\mathbb{R}\times S^2$ (with $S^2$ equipped with the round metric). Inspired by this isomorphism, it was shown in \cite{Saha:2023hsl} how the EM tensor Ward identities of a $1+2$D source-less CarrCFT on a Carrollian background with topology $\mathbb{R}\times S^2$ with assumed Weyl invariance can encode the $1+3$D bulk AFS leading \cite{Weinberg:1965nx,He:2014laa} and the subleading \cite{Cachazo:2014fwa,Kapec:2014opa} soft graviton theorems as well as the 2D Celestial CFT EM tensor \cite{Kapec:2016jld} and the (BMS$_4$) super-translation Kac-Moody \cite{Strominger:2013jfa} Ward identities.

\medskip

In addition to the leading \cite{Weinberg:1965nx} and subleading \cite{Cachazo:2014fwa} soft graviton theorems, another soft graviton theorem at the subsubleading order was found in \cite{Cachazo:2014fwa} explicitly for tree-level Einstein gravity in the $1+3$D bulk AFS. But, unlike the soft theorems at the two more leading orders, this subsubleading soft theorem is not universal \cite{Elvang:2016qvq,Laddha:2017ygw}. Equivalently in the Carrollian picture, the fact that the special Carrollian conformal fields $S^+_0$, $S^+_1$ and $T$, whose Ward identities contain the two universal AFS soft graviton theorems can be constructed purely out of the CarrCFT EM tensor \cite{Saha:2023hsl}, points towards an universality in the sense that every CarrCFT is expected to have an EM tensor. Consistently, the non-universal AFS subsubleading soft graviton theorems were not captured by those CarrCFT Ward identities.

\medskip

In this work, we aim to understand how the $1+3$D bulk AFS subsubleading soft graviton theorem \cite{Cachazo:2014fwa} can be encoded in the framework of the $1+2$D CarrCFT. Obviously, to provide a holographic description of a non-universal phenomenon in the bulk, the dual boundary theory also must possess some non-universal features. For this purpose, taking cue from the field contents of a 2D Euclidean CFT \cite{Zamolodchikov:1985wn} or a $1+1$D CarrCFT \cite{Bagchi:2023dzx} enjoying infinite additional symmetries, we postulate that, in a $1+2$D CarrCFT that carries the imprint of the positive-helicity subsubleading soft graviton theorem, there is a local quantum field $S^+_2$ in addition to the three universal local generators $S^+_0$, $S^+_1$ and $T$.

\medskip

To avoid the potentially problematic hologram of the bulk ambiguity associated with the double soft-limits of opposite helicities \cite{Klose:2015xoa}, we do not attempt to simultaneously describe the negative-helicity subsubleading soft theorem in this work. By ignoring one helicity, we are not inviting inconsistency per se --- we are focusing only on the holomorphic-sector of the $1+2$D CarrCFT; so, our holographic analysis will correspond only to the positive-helicity sectors of the bulk theories of gravitons. An exception is the MHV sector \cite{Banerjee:2020zlg} of the tree-level Einstein gravity, where the opposite helicity soft sectors decouple and our conclusions are rendered applicable to both sectors (with trivial modifications). 

\medskip

Following the methods, elaborated in \cite{Saha:2023hsl}, to completely determine the singular parts of the mutual operator product expansions (OPEs) of the generators $S^+_0$ and $S^+_1$, we derive the mutual OPEs of $S^+_2$ and the two aforementioned generators. This algorithm only makes use of the general Carrollian symmetry principles and the OPE commutativity property under the assumption that there are no time-independent (in the OPE limit) local fields with negative Carrollian scaling dimension in the CarrCFT. Similar symmetry arguments (together with similar assumptions) fixed the singularities of the mutual OPEs of the symmetry generators in the cases of 2D Euclidean CFT in \cite{Zamolodchikov:1985wn,Zamolodchikov:1989mz} and of $1+1$D CarrCFT in \cite{Saha:2022gjw,Bagchi:2023dzx}.  

\medskip

While trying to find the $S^+_2S^+_2$ OPE, we realize that for this OPE to be consistent, another local Carrollian conformal field $S^+_3$ must automatically appear in the theory. Extending the algorithm, we find in general that for the CarrCFT OPE $S^+_2S^+_{k\geq2}$ to be consistent, there must already exist an infinite tower of local fields $S_{k+1}^+(t,z,\bar{z})$. The general CarrCFT OPE $S^+_kS^+_l$ \eqref{37} is the main result of this work.

\medskip

From this general $S^+_k(\mathbf{x})S^+_l(\mathbf{x}_p)$ OPE \eqref{37}, we immediately recover the 2D Celestial conformal OPE of two conformally soft \cite{Donnay:2018neh,Puhm:2019zbl} primary gravitons $H^{1-k}(z,\bar{z})H^{1-l}(z_p,\bar{z}_p)$ of \cite{Guevara:2021abz}, just by inspection. This conformally soft graviton OPE in \cite{Guevara:2021abz} was directly obtained by taking the conformally soft limit of the general OPE between two general 2D Celestial conformal primary gravitons of arbitrary weights, derived in \cite{Pate:2019lpp}. But the derivation of the general graviton primary OPE in \cite{Pate:2019lpp} was explicitly for the tree-level (linearized) Einstein theory in the bulk AFS and required some hints from this specific bulk theory to fix the singular structure of the said OPE. In another method, this OPE was obtained via a Mellin transformation from the bulk graviton scattering amplitude in the collinear limit in \cite{Pate:2019lpp,Guevara:2021abz}. 

\medskip

It is to be emphasized that we have not obtained the general Celestial conformal primary graviton OPE of \cite{Pate:2019lpp} since that is a theory-specific result. Rather, we have given a completely holographic, Carrollian conformal symmetric derivation of the OPEs between the symmetry generators \cite{Guevara:2021abz} that is expected to contain the asymptotic symmetry algebra of any (quantum) gravity theory in the $1+3$D bulk AFS. Another important difference is that while the above mentioned Celestial CFT OPEs in \cite{Pate:2019lpp,Guevara:2021abz} are valid only at tree-level Einstein theory (but exact for quantum self-dual gravity \cite{Ball:2021tmb}), the Carrollian conformal derivation of the $S^+_kS^+_l$ OPE (under the aforementioned assumption) involves no (Carrollian) perturbation theory analysis; its starting point is the CarrCFT Ward identities derived in \cite{Saha:2023hsl} using a Carrollian path-integral formalism \cite{Chen:2023pqf}.

\medskip

Using the $1+2$D CarrCFT OPE $\longleftrightarrow$ commutation-relation prescription developed in \cite{Saha:2023hsl}, we find that the (local) symmetry algebra manifest in the CarrCFT OPE \eqref{37} is the Kac-Moody algebra of the wedge subalgebra \cite{Pope:1991ig} of the $w_{1+\infty}$ algebra \cite{Bakas:1989xu}, in perfect agreement with the symmetry algebra derived in \cite{Strominger:2021lvk} from the Celestial conformally soft primary graviton OPE of \cite{Guevara:2021abz}.

\medskip

Finally, we shed light on the direct connection between the existence of the infinite tower of Carrollian fields $S^+_k$ and an infinity of the soft graviton theorems discussed in \cite{Hamada:2018vrw,Li:2018gnc} in the context of the tree-level (linearized) Einstein theory. We find that the Ward identity of the Carrollian field $S^+_2$ does indeed encode up to the positive helicity subsubleading soft graviton theorems \cite{Cachazo:2014fwa}. While it is hinted that the Ward identities of the other fields $S^+_{k>2}$ contain the soft theorems in more subleading orders, it is also demonstrated that the fields beyond $S^+_2$ does not generate any new (independent) global symmetries of the theory. This is the CarrCFT analogue of the fact that the Celestial conformally soft gravitons $H^{k<-1}$ do not impose any new constraints on the $1+3$D bulk AFS $S$-matrices \cite{Guevara:2021abz}.

\medskip

The rest of the paper is organized as follows. In section \ref{s2}, we review the main results of \cite{Saha:2023hsl} on the universal features of a $1+2$D CarrCFT. We introduce the Carrollian conformal field $S^+_2$ whose Ward identity (supposedly) encodes the $1+3$D bulk AFS positive-helicity subsubleading soft graviton theorem, in section \ref{s3}. We state the assumptions on the field content of the CarrCFT in section \ref{s3.1} and find that for the $S^+_2S^+_2$ OPE to be consistent, a local field $S^+_3$ must already exist in the theory. Our purely symmetry-based algorithm to find the OPEs reveal the automatic existence of the infinite tower of Carrollian fields $S^+_{k\geq3}$ in section \ref{s4}. We obtain the general $S^+_kS^+_l$ OPE in section \ref{s4.3}, from which the quantum symmetry algebra is determined in section \ref{s5}. Finally, in section \ref{s6} we relate the infinite tower of fields $S^+_{k\geq2}$ with an infinite number of soft graviton theorems before concluding in section \ref{s7}.      

\medskip

\section{Review}\label{s2}
In \cite{Saha:2023hsl}, it was shown how the EM tensor Ward identities of an honest (i.e. source-less) Carrollian CFT on a $1+2$D flat Carrollian background (with topology $\mathbb{R}\times S^2$ and the assumption of the Weyl invariance) can be recast into the forms resembling to the $1+3$D bulk AFS leading and subleading conformally soft \cite{Donnay:2018neh,Puhm:2019zbl} graviton theorems \cite{Banerjee:2018fgd,Banerjee:2020zlg}. The $1+2$D Carrollian conformal fields $S^\pm_0$ and $S^\pm_1$ containing respectively the leading and the subleading conformally soft graviton primaries of the 2D Celestial CFT, as well as the fields $T$ and $\bar{T}$ that contain respectively the holomorphic and the anti-holomorphic components of the 2D Celestial CFT EM tensor \cite{Kapec:2016jld} were constructed purely out of the Carrollian EM tensor components $T^{\mu}_{\hspace{1.5mm}\nu}$. Below we note the generic Ward identities\footnote{All the correlators and OPEs considered in this work are implicitly covariant time-ordered (as defined in section $6.1.4.$ of \cite{Itzykson:1980rh}). Covariant time-ordering commutes with space-time differentiation and integration.} of the $1+2$D CarrCFT EM tensor, derived in \cite{Saha:2023hsl} using the Carrollian path-integral formalism \cite{Chen:2023pqf}:
\begin{align}
&\text{Translation:\hspace{2.5mm}}\partial_\mu\langle T^\mu_{\hspace{1.5mm}\nu}(\mathbf{x})X\rangle=-i\sum\limits_{p=1}^n\text{ }{{\partial_{\nu_p}}\langle X\rangle}\text{ }\delta(t-t_p)\delta^2(\vec{x}-\vec{x}_p)\label{eq:2}\\
&\text{Boost:\hspace{2.5mm}}\langle T^i_{\hspace{1.5mm}t}(\mathbf{x})X\rangle=-i\sum\limits_{p=1}^n\text{ }{(\bm{\xi_i})}_p\cdot{\langle X\rangle}\text{ }\delta(t-t_p)\delta^2(\vec{x}-\vec{x}_p)\label{eq:3}\\
&\text{Dilation:\hspace{2.5mm}}\langle T^\mu_{\hspace{1.5mm}\mu}(\mathbf{x})X\rangle=-i\sum\limits_{p=1}^n\text{ }{\Delta}_p{\langle X\rangle}\text{ }\delta(t-t_p)\delta^2(\vec{x}-\vec{x}_p)\label{eq:4}\\
&\text{Rotation:\hspace{2.5mm}}\langle T^z_{\hspace{1.5mm}z}(\mathbf{x})X\rangle-\langle T^{\bar{z}}_{\hspace{1.5mm}\bar{z}}(\mathbf{x})X\rangle=-i\sum\limits_{p=1}^n\text{ }s_p{\langle X\rangle}\text{ }\delta(t-t_p)\delta^2(\vec{x}-\vec{x}_p)\label{eq:5}
\end{align}
where $X$ is a string of Carrollian conformal primary multiplets; ${(\bm{\xi_i})}_p$ is the Carrollian boost (in the $i$-th direction) representation-matrix, $s_p$ is the spin (i.e. the eigenvalue of the spatial rotation) and $\Delta_p$ is the Carrollian conformal weight of the $p$-th primary field.

\medskip

\subsection{The fields $S^\pm_0$}
Subtraction of the spatial divergence of \eqref{eq:3} from \eqref{eq:2}$_{\nu=t}$ leads to:
\begin{align}
\partial_t\langle T^t_{\hspace{1.5mm}t}(t,\vec{x})X\rangle=-i\sum\limits_{p=1}^n\text{ }\delta(t-t_p)\left[\delta^2(\vec{x}-\vec{x}_p)\partial_{t_p}-\left(\vec{\bm{\xi}}_p\cdot\bm{\nabla}\right)\delta^2(\vec{x}-\vec{x}_p)\right]\langle X\rangle
\end{align}
Choosing the following initial condition:
\begin{align}
\langle T^t_{\hspace{1.5mm}t}(t\rightarrow-\infty,\vec{x})X\rangle=0\label{eq:6}
\end{align}
the solution to the above temporal partial differential equation is obtained as:
\begin{align}
\langle T^t_{\hspace{1.5mm}t}(t,\vec{x})X\rangle=-i\sum\limits_{p=1}^n\text{ }\theta(t-t_p)\left[\delta^2(\vec{x}-\vec{x}_p)\partial_{t_p}-\left(\vec{\bm{\xi}}_p\cdot\bm{\nabla}\right)\delta^2(\vec{x}-\vec{x}_p)\right]\langle X\rangle\label{eq:7}
\end{align} 

\medskip

The $S^2$ contact-term singularities in \eqref{eq:7} can be converted into pole singularities (but avoiding branch-cuts) if we note that:
\begin{align}
\langle {T}^t_{\hspace{1.5mm}t}(t,\vec{x})X\rangle&=-\frac{i}{\pi}\sum\limits_{p=1}^n\text{ }\theta(t-t_p)\text{ }{\bar{\partial}}^2\left[\frac{\bar{z}-\bar{z}_p}{z-z_p}\partial_{t_p}+\frac{\bar{z}-\bar{z}_p}{{(z-z_p)}^2}\bm{\xi}_p-\frac{\bar{{\bm{\xi}}}_p}{z-z_p}\right]\langle X\rangle\label{eq:8}\\
&=-\frac{i}{\pi}\sum\limits_{p=1}^n\text{ }\theta(t-t_p)\text{ }{{\partial}}^2\left[\frac{z-z_p}{\bar{z}-\bar{z}_p}\partial_{t_p}+\frac{z-z_p}{({\bar{z}-\bar{z}_p})^2}\bar{\bm{\xi}}_p-\frac{{{\bm{\xi}}}_p}{\bar{z}-\bar{z}_p}\right]\langle X\rangle\label{eq:9}
\end{align}
Inverting the ${\bar{\partial}}^2$ operator in \eqref{eq:8} and the $\partial^2$ operator in \eqref{eq:9}, the Carrollian fields $S^\pm_0$ were respectively defined in \cite{Saha:2023hsl} as:
\begin{align}
&S_0^+(t,z,\bar{z}):=\int\limits_{S^2} d^2{r}^\prime\text{ }\frac{\bar{z}-\bar{z}^\prime}{z-z^\prime}\text{ }{T}^t_{\hspace{1.5mm}t}(t,\vec{x}^\prime)\hspace{2.5mm}\Longrightarrow\hspace{2.5mm}\bar{\partial}^2S_0^+=\bar{\partial}P=\pi {T}^t_{\hspace{1.5mm}t}\label{1}\\
&S_0^-(t,z,\bar{z}):=\int\limits_{S^2} d^2{r}^\prime\text{ }\frac{z-z^\prime}{\bar{z}-\bar{z}^\prime}\text{ }{T}^t_{\hspace{1.5mm}t}(t,\vec{x}^\prime)\hspace{2.5mm}\Longrightarrow\hspace{2.5mm}{\partial}^2S_0^-={\partial}\bar{P}=\pi {T}^t_{\hspace{1.5mm}t}\label{2}
\end{align}
where the descendant fields $P=\bar{\partial}S_0^+$ and $\bar{P}=\partial S^-_0$ consist respectively of the modes generating the holomorphic and the anti-holomorphic super-translations. The scaling dimension $\Delta$ and the spin $m$ of the fields $S^\pm_0$ are $(\Delta,m)=(1,\pm 2)$. So, the defining relations \eqref{1} and \eqref{2} imply that $S^\pm_0$ are the 2D shadow-transformations (on $S^2$) of each other \cite{Pasterski:2017kqt} by construction. Since, a field and its shadow (being a highly non-local integral transformation) can not both be treated as local fields in a theory \cite{Banerjee:2022wht}, only one among $S^\pm_0$ is to be chosen as a local field while relegating the other merely to its non-local shadow.

\medskip

We treat $S^+_0$ as the local field and $S^-_0$ as its shadow in this work. This corresponds to investigating the holomorphic sector of the $1+2$D Carrollian CFT. The $S^+_0$ Ward identity, for a string $X$ of $n$ mutually local Carrollian conformal primary multiplet fields $\left\{\Phi_p(t_p,z_p,\bar{z}_p)\right\}$, reads \cite{Saha:2023hsl}:
\begin{align}
&\langle S^+_0(t,z,\bar{z})X\rangle=-i\sum\limits_{p=1}^n\text{ }\theta(t-t_p)\text{ }\left\{\frac{\bar{z}-\bar{z}_p}{z-z_p}\partial_{t_p}+\frac{\bar{z}-\bar{z}_p}{{(z-z_p)}^2}\bm{\xi}_p-\frac{\bar{{\bm{\xi}}}_p}{z-z_p}\right\}\langle X\rangle\label{3}\\
\Longrightarrow\hspace{2.5mm}&\langle\bar{\partial}^2 S^+_0(t,z,\bar{z})X\rangle=-i\sum\limits_{p=1}^n\text{ }\theta(t-t_p)\text{ }\left[\text{contact terms on $S^2$}\right]\nonumber\\
\text{and}\hspace{2.5mm}&{\partial_t}\langle S^+_0(t,z,\bar{z})X\rangle=\left[\text{temporal contact terms}\right]\nonumber
\end{align}
where $\bm{\xi},\bar{\bm{{\xi}}}:=\bm{\xi}_x\pm i\bm{\xi}_y$ denote the matrix-representation of the classical Carrollian boost under which a Carrollian multiplet $\Phi$ transforms\footnote{See \cite{Chen:2021xkw} for a more general representation theory of the global Carrollian conformal algebra.}. 

\medskip

The above correlator was derived from the (Carrollian) super-translation Ward identity in \cite{Saha:2023hsl}. The temporal step-function appearing in this Carrollian correlator captures the essence of the $1+3$D bulk AFS super-translation memory effect \cite{Strominger:2014pwa,Strominger:2017zoo}. Consistently, temporal-Fourier transforming \eqref{3} to (positive) $\omega$-space and then making the $S^+_0$ field energetically soft \cite{Strominger:2014pwa}, one recovers the Weinberg (leading) positive-helicity soft graviton theorem \cite{Weinberg:1965nx,He:2014laa} as the residue of the leading $\frac{1}{\omega}$ pole when all of the primaries in $X$ have $\bm{\xi}=\bar{\bm{{\xi}}}=0$ \cite{Saha:2023hsl}. Thus, explicitly at the $t\rightarrow\infty$ limit, the Ward identity \eqref{3} is same as the positive-helicity leading conformally soft graviton theorem \cite{Banerjee:2018fgd} when all $\bm{\xi}_p=\bar{\bm{{\xi}}}_p=0$. 

\medskip

The $S^2$ stereographic coordinates $z$ and $\bar{z}$ are now treated as independent variables \cite{Fotopoulos:2019tpe} so that terms like $\frac{\left(\bar{z}-\bar{z}_p\right)^r}{\left(z-z_p\right)^s}$ with $r\geq0,s\geq1$ have (meromorphic) pole singularity (avoiding phase ambiguity when $r=s$). Together with the form of the Ward identity \eqref{3}, this suggests that, inside the correlator, $S^+_0$ can be decomposed as \cite{Banerjee:2020zlg}:
\begin{align}
&S^+_0(t,z,\bar{z})=\bar{z}P_{-1}(t,z,\bar{z})-P_0(t,z,\bar{z})\label{4}\\
\text{with}\hspace{5mm}&\langle P_{-1}(t,z,\bar{z})X\rangle=-i\sum\limits_{p=1}^n\text{ }\theta(t-t_p)\left(\frac{\partial_{t_p}}{z-z_p}+\frac{\bm{\xi}_p}{{(z-z_p)}^2}\right)\langle X\rangle\nonumber\\
\text{and}\hspace{5mm}&\langle P_0(t,z,\bar{z})X\rangle=-i\sum\limits_{p=1}^n\text{ }\theta(t-t_p)\left(\frac{\bar{z}_p\partial_{t_p}+\bar{{\bm{\xi}}}_p}{z-z_p}+\frac{\bar{z}_p\bm{\xi}_p}{{(z-z_p)}^2}\right)\langle X\rangle\nonumber\\
\Longrightarrow\hspace{5mm}&\langle\bar{\partial} P_i(t,z,\bar{z})X\rangle=-i\sum\limits_{p=1}^n\text{ }\theta(t-t_p)\text{ }\left[\text{contact terms on $S^2$}\right]\nonumber
\end{align}
These relations reminisce holomorphic Kac-Moody like Ward identities in a 2D Euclidean CFT. Clearly, $P_i$ and $S^+_0$ have the same holomorphic weight $h=\frac{3}{2}$.

\medskip

Let us next discuss on the Carrollian conformal OPEs. As explained in \cite{Saha:2023hsl,Saha:2022gjw}, it is convenient to convert the temporal step-function appearing in the correlators like \eqref{3} into a $j\epsilon$-prescription for this purpose, with $j$ being a second complex unit. The starting point is to hyper-complexify the $(t,z,\bar{z})$ coordinates as below:
\begin{align*}
\hat{z}:=z+jt\hspace{2.5mm};\hspace{2.5mm}\hat{\bar{z}}:=\bar{z}+jt\hspace{2.5mm};\hspace{2.5mm}\hat{t}:=t
\end{align*}
While $z$ is a complex number on the $x-y$ plane, $\hat{z}$ can be thought of as a complex number on a $y=ax+b$ plane of the 3D $t-x-y$ space. $t>0$ is the upper-half of this plane.

\medskip

It can be shown that all of $\frac{\partial\hat{z}}{\partial\hat{t}},\frac{\partial\hat{\bar{z}}}{\partial\hat{t}},\frac{\partial\hat{\bar{z}}}{\partial\hat{z}}$ vanish. So, $\hat{t},\hat{z},\hat{\bar{z}}$ can be treated as independent variables. Thus, in most cases, we choose the point of insertion of a Carrollian field to be at $(\hat{t},\hat{z},\hat{\bar{z}})=(t,z,\bar{z})$. E.g. the $j\epsilon$-form of the Ward identity \eqref{3} is (with $\Delta\tilde{z}_p:=z-z_p-j\epsilon(t-t_p)$ ):
\begin{align*}
i\langle S^+_0(t,{z},{\bar{z}})X\rangle=\lim\limits_{\epsilon\rightarrow0^+}\sum\limits_{p=1}^n\left\{\frac{\bar{z}-\bar{z}_p}{(\Delta\tilde{z}_p)}\partial_{t_p}+\frac{\bar{z}-\bar{z}_p}{{(\Delta\tilde{z}_p)}^2}\bm{\xi}_p-\frac{\bar{{\bm{\xi}}}_p}{(\Delta\tilde{z}_p)}\right\}\langle X\rangle
\end{align*}
The main application of this $j\epsilon$-prescription is to establish the relation between the CarrCFT OPEs and the corresponding commutation relations while allowing for a straightforward utilization of the OPE commutativity property.

\medskip

It is important to remember that $\frac{1}{\Delta\tilde{z}_p}$ reduces to $\frac{1}{z-z_p}$ only when $t-t_p>0$ and to $0$ when $t-t_p<0$ in the sense of distributions \cite{Saha:2022gjw} encoding the property of the temporal step-function. Thus, $\frac{1}{\Delta\tilde{z}_p}\equiv \frac{1}{z-{z}_p}$ when $t\rightarrow\infty$, and $\frac{1}{\Delta\tilde{z}_p}\equiv0$ when $t\rightarrow-\infty$, thus providing a justification to the initial condition \eqref{eq:6}.

\medskip

The OPE\footnote{The two Carrollian operators whose product is to be expanded are inserted at different spatial-positions as well as at different times \cite{Saha:2022gjw}.} of $S^+_0$ with a general (non-primary) Carrollian conformal field $\Phi$ (that is mutually local with $S^+_0$) is noted below \cite{Saha:2023hsl}:
\begin{align}
iS^+_0(t,z,\bar{z})\Phi(\mathbf{x}_p)\sim\lim\limits_{\epsilon\rightarrow0^+}&\left[(\bar{z}-\bar{z}_p)\left(\sum\limits_{n\geq0}^J\frac{\left(P_{n,-1}\Phi\right)}{{(\Delta\tilde{z}_p)}^{n+2}}+\frac{\partial_{t_p}\Phi}{(\Delta\tilde{z}_p)}\right)-\sum\limits_{n\geq-1}^K\frac{\left(P_{n,0}\Phi\right)}{{(\Delta\tilde{z}_p)}^{n+2}}\right](\mathbf{x}_p)\label{5}\\
\Longrightarrow\hspace{5mm}&\bar{\partial}^2 S^+_0(t,z,\bar{z})\Phi(\mathbf{x}_p)\sim0\label{6}\\
\text{and}\hspace{5mm}&\partial_t S^+_0(t,z,\bar{z})\Phi(\mathbf{x}_p)\sim0\label{7}
\end{align}
where $\sim$ denotes `modulo terms holomorphic (regular) in $\Delta\tilde{z}_p$ '. This OPE is a Laurent series in the holomorphic variable $z$ (or $\tilde{z}=z-j\epsilon t$) but is (anti-)holomorphic (i.e. a Taylor series) in $\bar{z}$. Specially, a Carrollian conformal primary multiplet field $\Phi$ satisfies:
\begin{align*}
\left(P_{n+1,-1}\Phi\right)=0=\left(P_{n,0}\Phi\right)\text{\hspace{5mm} for $n\geq0$}
\end{align*}
while a Carrollian conformal quasi-primary (i.e. an only $\text{ISL}(2,\mathbb{C})$ covariant) multiplet only needs $\left(P_{0,0}\Phi\right)=0$.

\medskip  

\subsection{The fields $S^\pm_1$}
Next, we combine the CarrCFT EM tensor Ward identities \eqref{eq:4}, \eqref{eq:5} and \eqref{eq:7} into the following form:
\begin{align*}
&\langle T^z_{\hspace{1.5mm}z}(\mathbf{x})X\rangle+\frac{1}{2}\langle T^t_{\hspace{1.5mm}t}(\mathbf{x})X\rangle=-i\sum\limits_{p=1}^n\text{ }{h}_p{\langle X\rangle}\text{ }\delta(t-t_p)\delta^2(\vec{x}-\vec{x}_p)\\
\Rightarrow\text{ }& i\langle T^z_{\hspace{1.5mm}z}(\mathbf{x})X\rangle=\sum\limits_{p=1}^n\left[\delta(t-t_p)\delta^2(\vec{x}-\vec{x}_p)h_p-\frac{\theta(t-t_p)}{2}\left\{\delta^2(\vec{x}-\vec{x}_p)\partial_{t_p}-\left(\vec{\bm{\xi}}_p\cdot\bm{\nabla}\right)\delta^2(\vec{x}-\vec{x}_p)\right\}\right]{\langle X\rangle}
\end{align*}
Thus, subtraction of $\partial_z\langle T^z_{\hspace{1.5mm}z}(\mathbf{x})X\rangle$ from \eqref{eq:2}$_{\nu=z}$ results into:
\begin{align*}
\partial_{\bar{z}}\langle T^{\bar{z}}_{\hspace{1.5mm}z}(\mathbf{x})X\rangle+\partial_t\langle T^t_{\hspace{1.5mm}z}(\mathbf{x})X\rangle&=-i\sum\limits_{p=1}^n\left[\delta(t-t_p)\left\{\delta^2(\vec{x}-\vec{x}_p)\partial_{z_p}-h_p\partial_{z}\delta^2(\vec{x}-\vec{x}_p)\right\}\right.\nonumber\\
&\left.+\frac{\theta(t-t_p)}{2}\partial_z\left\{\delta^2(\vec{x}-\vec{x}_p)\partial_{t_p}-\left(\vec{\bm{\xi}}_p\cdot\bm{\nabla}\right)\delta^2(\vec{x}-\vec{x}_p)\right\}\right]{\langle X\rangle}
\end{align*}
Choosing an initial condition similar to \eqref{eq:6}, one obtains the following solution to the above temporal partial differential equation:
\begin{align}
\langle T^t_{\hspace{1.5mm}z}(\mathbf{x})X\rangle+\int\limits^t_{-\infty}dt^\prime\text{ }\partial_{\bar{z}}\langle T^{\bar{z}}_{\hspace{1.5mm}z}(t^\prime,\vec{x})X\rangle=&-i\sum\limits_{p=1}^n\theta(t-t_p)\left[\delta^2(\vec{x}-\vec{x}_p)\partial_{z_p}-h_p\partial_{z}\delta^2(\vec{x}-\vec{x}_p)\right.\nonumber\\
&\left.+\frac{t-t_p}{2}\partial_z\left\{\delta^2(\vec{x}-\vec{x}_p)\partial_{t_p}-\left(\vec{\bm{\xi}}_p\cdot\bm{\nabla}\right)\delta^2(\vec{x}-\vec{x}_p)\right\}\right]{\langle X\rangle}\label{a13}
\end{align}
Its `complex-conjugated' version can be obviously derived in an exactly similar way.

\medskip

To convert the $S^2$ contact terms in \eqref{a13} (or, its complex-conjugate) into pole singularities, we extract a $\partial^3$ (or, $\bar{\partial}^3$) -derivative from the R.H.S.; inverting these derivatives, the Carrollian conformal fields $S^\pm_1$ that contain the sub-leading conformally soft gravitons \cite{Donnay:2018neh,Puhm:2019zbl} were defined in \cite{Saha:2023hsl} as:
\begin{align}
&S_1^-(t,z,\bar{z})=\int\limits_{S^2} d^2{r}^\prime\text{ }\frac{{(z-z^\prime)}^2}{\bar{z}-\bar{z}^\prime}\left[T^t_{\hspace{1.5mm}z}(t,\vec{x}^\prime)+\int\limits_{-\infty}^tdt^\prime\partial_{\bar{z}^\prime}T^{\bar{z}}_{\hspace{1.5mm}z}(t^\prime,\vec{x}^\prime)\right]\\
&S_1^+(t,z,\bar{z})=\int\limits_{S^2} d^2{r}^\prime\text{ }\frac{{(\bar{z}-\bar{z}^\prime)}^2}{{z}-{z}^\prime}\left[T^t_{\hspace{1.5mm}\bar{z}}(t,\vec{x}^\prime)+\int\limits_{-\infty}^tdt^\prime\partial_{{z}^\prime} T^{{z}}_{\hspace{1.5mm}\bar{z}}(t^\prime,\vec{x}^\prime)\right]
\end{align}
The dimensions of the fields $S^\pm_1$ are $(\Delta,m)=(0,\pm2)$.

\medskip

The $S^+_1$ Ward identity with a string of mutually local Carrollian conformal primaries was obtained as \cite{Saha:2023hsl}:
\begin{align}
i&\langle S_1^+(t,z,\bar{z}) X\rangle=\sum\limits_{p=1}^n\theta(t-t_p)\left[\frac{(\bar{z}-\bar{z}_p)^2}{z-{z}_p}\partial_{\bar{z}_p}-2\bar{h}_p\frac{\bar{z}-\bar{z}_p}{z-{z}_p}\right.\nonumber\\
&\hspace{44mm}\left.+(t-t_p)\left(\frac{\bar{z}-\bar{z}_p}{z-{z}_p}\partial_{t_p}+\frac{\bar{z}-\bar{z}_p}{{(z-{z}_p)}^2}\bm{\xi}_p-\frac{\bar{{\bm{\xi}}}_p}{z-{z}_p}\right)\right]\langle X\rangle\label{10}\\
\Longrightarrow\hspace{5mm}&\langle\bar{\partial}^3 S^+_1(t,z,\bar{z})X\rangle=-i\sum\limits_{p=1}^n\text{ }\theta(t-t_p)\text{ }\left[\text{contact terms on $S^2$}\right]\nonumber\\
\text{and}\hspace{5mm}&{\partial_t}\langle S^+_1(t,z,\bar{z})X\rangle-\langle S^+_0(t,z,\bar{z})X\rangle=\left[\text{temporal contact terms}\right]\label{11}
\end{align}
that for all $\bm{\xi}_p=\bar{\bm{{\xi}}}_p=0$ resembles (in the limit $t\rightarrow\infty$) the positive-helicity subleading conformally soft graviton theorem as presented (but very differently interpreted) in \cite{Banerjee:2020zlg}. More appropriately, the Ward identity \eqref{10} is the $1+2$D Carrollian conformal manifestation of the 2D Celestial subleading conformally soft graviton theorem.

\medskip

This Ward identity was derived from the (Carrollian) super-rotation Ward identity in \cite{Saha:2023hsl}. There, it was also shown that, upon temporal-Fourier transforming \eqref{10} and then taking the $\omega\rightarrow0^+$ limit only for $S^+_1$ \cite{Pasterski:2015tva}, one obtains a Laurent expansion around $\omega=0$, the coefficient of the leading $\frac{1}{\omega^2}$ pole of which is the Weinberg positive-helicity soft-graviton theorem \cite{Weinberg:1965nx,He:2014laa} while the subleading $\frac{1}{\omega}$ pole's coefficient is recognized to be the Cachazo-Strominger subleading positive-helicity soft-graviton theorem \cite{Cachazo:2014fwa,Kapec:2014opa} when each of the primaries in $X$ has $\bm{\xi}=\bar{\bm{{\xi}}}=0$ and (Carrollian) scaling dimension $\Delta=1$. Therefore, the temporal step-function in \eqref{10} is the Carrollian manifestation of the super-rotation memory effect \cite{Pasterski:2015tva,Strominger:2017zoo}. 

\medskip

The last condition on the scaling dimension of a Carrollian conformal primary whose temporal-Fourier transformation can correspond to a $1+3$D bulk AFS null momentum-space field (see also \cite{Bekaert:2022ipg,Liu:2022mne}) describing a mass-less external hard scattering particle, was discovered in \cite{Donnay:2022aba,Donnay:2022wvx} by analyzing the radiative fall-off conditions of the bulk mass-less fields. The higher dimensional counterpart of this condition was obtained more recently in \cite{Nguyen:2023vfz}. 

\medskip
  
\eqref{11} and \eqref{5} motivate us to re-express the $S^+_1$ field inside the correlator as below \cite{Saha:2023hsl}:
\begin{align}
&S^+_1(t,z,\bar{z})=S^+_{1e}(t,z,\bar{z})+tS^+_0(t,z,\bar{z})\label{12}\\
\Longrightarrow\hspace{5mm}&{\partial_t}\langle S^+_{1e}(t,z,\bar{z})X\rangle=\left[\text{temporal contact terms}\right]\nonumber\\
\text{and}\hspace{5mm}&\langle\bar{\partial}^3 S^+_{1e}(t,z,\bar{z})X\rangle=-i\sum\limits_{p=1}^n\text{ }\theta(t-t_p)\text{ }\left[\text{contact terms on $S^2$}\right]\nonumber
\end{align}
It should be noted that $S^+_{1e}$ is not a local Carrollian field but merely a collection of the modes not appearing in the $S^+_0$ field. These modes generate the following local infinitesimal Carrollian diffeomorphisms:
\begin{align*}
z\rightarrow z^\prime=z\hspace{2.5mm};\hspace{2.5mm}\bar{z}\rightarrow \bar{z}^\prime=\bar{z}+\varepsilon\bar{z}^{q+1}f(z)\hspace{2.5mm};\hspace{2.5mm}t\rightarrow t^\prime=t+\varepsilon \frac{t}{2}(q+1)\bar{z}^{q} f(z)
\end{align*}
with $f(z)$ being a meromorphic function and $q=0,\pm1$. It is the `Ward identity' $\langle S^+_{1e}(t\rightarrow\infty,z,\bar{z})X\rangle$ that directly gives rise to the Cachazo-Strominger subleading energetically soft graviton theorem \cite{Cachazo:2014fwa,Kapec:2014opa}.

\medskip

Since, $z$ and $\bar{z}$ are treated independently, the form of \eqref{10} allows us to decompose $S^+_{1e}$ inside a correlator as \cite{Banerjee:2020zlg,Polyakov:1987zb}:
\begin{align}
&S^+_{1e}(t,z,\bar{z})=j^{(+)}_e(t,z,\bar{z})-2\bar{z}j^{(0)}_e(t,z,\bar{z})+\bar{z}^2j^{(-)}_e(t,z,\bar{z})\label{18}\\
\Longrightarrow\hspace{5mm}&\langle\bar{\partial} j^a_e(t,z,\bar{z})X\rangle=-i\sum\limits_{p=1}^n\text{ }\theta(t-t_p)\text{ }\left[\text{contact terms on $S^2$}\right]\nonumber
\end{align}
Since, all three $j^a_e$ have holomorphic weight $h=1$ like $S^+_1$, their Ward identities are effectively same as the holomorphic Kac-Moody Ward identities in a usual 2D CFT.

\medskip

Finally, we note down the general OPE of the $S^+_1$ field in the $j\epsilon$-form \cite{Saha:2023hsl}:
\begin{align}
&i\left\{S_1^+-({t-t_p})S^+_0\right\}(t,z,\bar{z})\Phi(\mathbf{x}_p)\sim\lim\limits_{\epsilon\rightarrow0^+}\left[(\bar{z}-\bar{z}_p)^2\left(\sum\limits_{n\geq1}^L\frac{\left(j^{(-)}_{n}\Phi\right)}{{(\Delta\tilde{z}_p)}^{n+1}}+\frac{\partial_{\bar{z}_p}\Phi}{\Delta\tilde{z}_p}\right)+\sum\limits_{n\geq0}^M\frac{\left(j^{(+)}_{n}\Phi\right)}{{(\Delta\tilde{z}_p)}^{n+1}}\right.\nonumber\\
&\hspace{63mm}\left.-2(\bar{z}-\bar{z}_p)\left(\sum\limits_{n\geq1}^N\frac{\left(j^{(0)}_{n}\Phi\right)}{{(\Delta\tilde{z}_p)}^{n+1}}+\frac{\bar{h}_p\Phi}{\Delta\tilde{z}_p}\right)\right](\mathbf{x}_p)\label{14}\\
&\hspace{18mm}\Longrightarrow\hspace{5mm}\bar{\partial}^3 S^+_1(t,z,\bar{z})\Phi(\mathbf{x}_p)\sim0\label{15}\\
&\hspace{18mm}\text{and}\hspace{5mm}\left(\partial_t S^+_1-S^+_0\right)(t,z,\bar{z})\Phi(\mathbf{x}_p)\sim0\label{16}
\end{align}
For a Carrollian conformal primary multiplet $\Phi$, we have:
\begin{align*}
\left(j^{(-)}_{n}\Phi\right)=\left(j^{(0)}_{n}\Phi\right)=\left(j^{(+)}_{n-1}\Phi\right)=0\text{\hspace{5mm} for $n\geq1$}
\end{align*}
whereas an $\text{SL}(2,\mathbb{C})$ or Lorentz covariant quasi-primary must satisfy: $\left(j^{(+)}_{0}\Phi\right)=0$.

\medskip

Obviously, the corresponding correlators and OPEs involving the $S^-_1$ field (replacing $S^+_1$) are just the complex conjugations $(z\rightarrow\bar{z},\bar{h}\rightarrow h,\bm{\xi}\rightarrow\bm{\bar{\xi}})$ of the above mentioned equations. But as shown in \cite{Saha:2023hsl}, following \cite{Banerjee:2022wht}, the fields $S^+_0$ and $S^-_1$ can not be simultaneously treated as local fields i.e. they are not mutually local. The reason is that while $\left(\bar{\partial}^2 S^+_0\right)S^-_1\sim0$ respects \eqref{6}, $S^-_1\bar{\partial}^2 S^+_0$ contains anti-meromorphic pole singularities, thus violating the OPE commutativity property. This is a Carrollian manifestation of the ordering ambiguity in the double soft limit involving two opposite helicity particles \cite{Klose:2015xoa}. On the other hand, the OPE of $S^+_0$ with $S^+_1$ does not suffer from this problem; hence, $S^+_0$ and $S^+_1$ can both be simultaneously taken as local fields that we do.

\medskip

As we shall see, the OPE conditions like \eqref{6}, \eqref{7}, \eqref{15} and \eqref{16} will play very crucial roles in this work.

\medskip

\subsection{The fields $T$ and $\bar{T}$}
All the $S^2$ contact terms in the Ward identity \eqref{a13} (or, its conjugate version) can also be converted into pole singularities by extracting from the R.H.S. a $\bar{\partial}$ (or, $\partial$) -derivative. The Carrollian fields $T$ and $\bar{T}$ were then constructed in \cite{Saha:2023hsl} by inverting these derivatives as:
\begin{align}
&T(t,z,\bar{z})=\int\limits_{S^2} d^2{r}^\prime\frac{T^t_{\hspace{1.5mm}z}(t,\vec{x}^\prime)}{z-z^\prime}+\pi\int\limits_{-\infty}^tdt^\prime T^{\bar{z}}_{\hspace{1.5mm}z}(t^\prime,\vec{x})\hspace{2.5mm}\Longrightarrow\hspace{2.5mm}2\bar{\partial}T=\partial^3S^-_1\\
&\bar{T}(t,z,\bar{z})=\int\limits_{S^2} d^2{r}^\prime\frac{T^t_{\hspace{1.5mm}\bar{z}}(t,\vec{x}^\prime)}{\bar{z}-\bar{z}^\prime}+\pi\int\limits_{-\infty}^tdt^\prime T^{{z}}_{\hspace{1.5mm}\bar{z}}(t^\prime,\vec{x})\hspace{2.5mm}\Longrightarrow\hspace{2.5mm}2{\partial}\bar{T}=\bar{\partial}^3S^+_1
\end{align}
Since, the field $T$ has dimensions $(\Delta,m)=(2,2)$ and $\bar{T}$ has $(\Delta,m)=(2,-2)$, the above relations, together with the fact that $S^\pm_1$ have $(\Delta,m)=(0,\pm2)$, imply that $(S^+_1,\bar{T})$ and $(S^-_1,T)$ automatically are two shadow pairs (on $S^2$). Since, we have chosen $S^+_1$ as a local field, $\bar{T}$ now has to be treated as its non-local shadow.

\medskip

With mutually local primaries, the $T$ Ward identity was derived from the super-rotation Ward identity to be \cite{Saha:2023hsl}:
\begin{align}
&i\langle T(t,z,\bar{z})X\rangle\nonumber\\
=&\sum\limits_{p=1}^n\theta(t-t_p)\left[\frac{h_p}{(z-z_p)^2}+\frac{\partial_{z_p}}{z-z_p}-\frac{t-t_p}{2}\left\{\frac{\partial_{t_p}}{(z-z_p)^2}+\frac{2\bm{\xi}_p}{{(z-z_p)}^3}+\pi\bar{\bm{\xi}}_p\partial_z\delta^2(\vec{x}-\vec{x}_p)\right\}\right]\langle X\rangle\\
&\text{with\hspace{2.5mm}}\langle \bar{\partial} T(t,z,\bar{z})X\rangle=-i\sum\limits_{p=1}^n\theta(t-t_p)\left[\text{contact terms on $S^2$}\right]\nonumber\\
&\text{and\hspace{2.5mm}}\partial_t\langle T(t,z,\bar{z})X\rangle-\frac{1}{2}\langle \partial\bar{\partial} S^+_0(t,z,\bar{z})X\rangle=[\text{temporal contact terms}]\nonumber
\end{align} 
Inspired by the last relation, we decompose the $T$ field inside the correlator as \cite{Saha:2023hsl}:
\begin{align}
&T(t,z,\bar{z})=\frac{t}{2}\partial\bar{\partial} S^+_0(t,z,\bar{z})+T_e(t,z,\bar{z})\label{46}\\
\Longrightarrow\hspace{2.5mm}&{\partial}_t\langle T_{e}(t,z,\bar{z}) X\rangle=[\text{temporal contact terms}]\nonumber
\end{align}
where $T_e$ is not a local Carrollian field but contains the modes generating the holomorphic super-rotations. It is the object $T_e(t\rightarrow\infty,z,\bar{z})$ that corresponds to the 2D Celestial stress-tensor \cite{Kapec:2016jld}; by construction, it is the 2D shadow transformation of the negative-helicity energetically soft graviton $S^-_{1e}(t\rightarrow\infty,z,\bar{z})$. The $\langle T_{e}(t,z,\bar{z}) X\rangle$ `Ward identity' is the Carrollian conformal analogue of that of the 2D Celestial holomorphic stress-tensor \cite{Fotopoulos:2019vac,Banerjee:2020kaa} when all $\bm{\xi}_p=\bar{\bm{{\xi}}}_p=0$.

\medskip

The generic OPE involving the $T$ field is given in the $j\epsilon$-form as \cite{Saha:2023hsl}:
\begin{align}
&i\left(T-\frac{t-t_p}{2}\text{ }\partial\bar{\partial}S^+_0\right)(t,z,\bar{z})\Phi(\mathbf{x}_p)\sim\lim\limits_{\epsilon\rightarrow0^+}\left[\sum\limits_{n\geq1}^P\frac{\left(L_{n}\Phi\right)}{{(\Delta\tilde{z}_p)}^{n+2}}+\frac{h_p\Phi}{(\Delta\tilde{z}_p)^2}+\frac{\partial_{z_p}\Phi}{\Delta\tilde{z}_p}\right](\mathbf{x}_p)\label{8}\\
\Longrightarrow\hspace{5mm}&\bar{\partial} T(t,z,\bar{z})\Phi(\mathbf{x}_p)\sim0\hspace{5mm}\text{and}\hspace{5mm}\left(\partial_t T-\frac{1}{2}\partial\bar{\partial} S^+_0\right)(t,z,\bar{z})\Phi(\mathbf{x}_p)\sim0\label{44}
\end{align}
For a Carrollian conformal primary, $\left(L_{n}\Phi\right)=0$ for $n\geq1$. An $\text{SL}(2,\mathbb{C})$ quasi-primary on the other hand needs to satisfy only $\left(L_{1}\Phi\right)=0$.

\medskip

In this work, we shall simultaneously treat the three Carrollian fields $S^+_0$, $S^+_1$ and $T$ as $\text{SL}(2,\mathbb{C})$ quasi-primary (non-descendant) local fields, as was shown to be allowed in \cite{Saha:2023hsl}.

\medskip

\section{The Carrollian Conformal Field $S^+_2$}\label{s3}
The features reviewed in the previous section are the generic properties of any (Weyl invariant) $1+2$D Carrollian CFT on flat (Carrollian) background. The Carrollian conformal Ward identities that were shown \cite{Saha:2023hsl,Donnay:2022aba,Donnay:2022wvx} to contain the equivalent information as the bulk AFS leading \cite{Weinberg:1965nx} and subleading \cite{Cachazo:2014fwa} soft graviton theorems were obtained as a consequence of only the Poincar\'e, the super-translation and the Weyl invariance in \cite{Saha:2023hsl}. In view of the putative AFS/CarrCFT duality \cite{Duval:2014uva}, this is in perfect agreement with the conclusions of \cite{Laddha:2017ygw} that in a generic theory of quantum gravity, only the leading and the subleading soft graviton theorems are universal. Consistently in \cite{deGioia:2023cbd}, only these two soft graviton theorems were reached via a `large AdS radius' limit from the AdS$_4$/CFT$_3$ correspondence. The non-universality of the subsubleading soft graviton \cite{Cachazo:2014fwa} theorem was discussed in \cite{Laddha:2017ygw,Elvang:2016qvq}.

\medskip

To probe the non-universal subsubleading soft graviton theorem that occurs, e.g. at the tree-level (linearized) Einstein gravity \cite{Cachazo:2014fwa}, it then naturally appears that additional Carrollian conformal fields beyond the usual Carrollian generators $S^\pm_0$, $S^\pm_1$, $T$ and $\bar{T}$ are required to be present in the CarrCFT. This situation is similar in spirit to those considered in \cite{Zamolodchikov:1985wn,Bagchi:2023dzx} where, besides the conformal EM tensor, extra symmetry generators were postulated to exist in the 2D theory.

\medskip

As we reviewed, the Carrollian fields $S^\pm_1$ encode both the $\pm$ve helicity leading and the subleading energetically soft graviton theorems while $S^\pm_0$ account for only the leading ones. The OPEs of the same spin fields among them are related by e.g. \eqref{16} while the $S^\pm_0$ OPEs satisfy e.g. \eqref{7}. Moreover, it is the temporal step-function (and its time-integrals) appearing in the Ward identities, from which the energetically-soft pole structures are arising. Inspired by these observations, \textit{we assume that, in the theory, there exists a Carrollian conformal field $S^+_2$ that, in the OPE limit, satisfies}:
\begin{align}
\left(\partial_t S^+_2-S^+_1\right)(t,z,\bar{z})\Phi(\mathbf{x}_p)\sim0\label{9}
\end{align}
Since $S^+_2$ is postulated to be a Carrollian field, it must contain some new modes. In view of \eqref{12}, this suggests that inside a correlator, $S^+_2$ can be decomposed as:
\begin{align}
&S^+_2(t,z,\bar{z})=S^+_{2e}(t,z,\bar{z})+tS^+_{1e}(t,z,\bar{z})+\frac{t^2}{2}S^+_0(t,z,\bar{z})\label{17}\\
\Longrightarrow\hspace{5mm}&{\partial_t}\langle S^+_{2e}(t,z,\bar{z})X\rangle=\left[\text{temporal contact terms}\right]\nonumber
\end{align}
where the $S^+_{2e}$ part consists of the new modes. Thus, the dimensions of the field $S^+_2$ are $(\Delta,m)=(-1,2)$.

\medskip

Clearly, an analogously introduced Carrollian field $S^-_2$ with spin $m=-2$ can not be treated as a mutually local field with $S^+_0$, $S^+_1$ and $T$, following the argument presented in \cite{Saha:2023hsl,Banerjee:2022wht}. While its shadow or light transformations (on $S^2$) may be fine in this regard, we leave this possibility for a future work. Thus, in this work, we refrain from introducing the $S^-_2$ field or its integral transformations.

\medskip

It is also evident that the shadow or the light transformations of the $S^+_2$ field can not be mutually local with $S^+_0$, $S^+_1$ and $T$. So, we would like $S^+_2$ itself to fit in as a local field in the holomorphic sector of the $1+2$D CarrCFT. It will be possible only if the $S^+_2\Phi$ OPEs have the similar singularity structure as those of the above three generators, i.e. having meromorphic pole singularities while being anti-analytic. We shall proceed by assuming this to be true.

\medskip

\subsection{The OPEs of $S^+_2$ with the generators}\label{s3.1}
In \cite{Saha:2023hsl}, the singular parts of mutual (self and cross) OPEs between the three generators $S^+_0$, $S^+_1$ and $T$ were completely determined by demanding the OPE commutativity property to hold, after making appropriate ansatz respecting the general forms \eqref{5}, \eqref{14} and \eqref{8} and truncating those ansatz by assuming that:
\begin{enumerate}
\item no local field in the theory has negative scaling dimension $\Delta<0$, following the 2D Euclidean \cite{Zamolodchikov:1985wn,Zamolodchikov:1989mz} and Carrollian \cite{Saha:2022gjw,Bagchi:2023dzx} CFT cases.
\item the fields $S^+_0$ and $S^+_1$ are Lorentz quasi-primaries, following the Celestial CFT case \cite{Fotopoulos:2019tpe}.
\end{enumerate}
The first assumption is clearly not respected by the field $S^+_2$ with $\Delta=-1$. So, it needs to be relaxed into a weaker one that is stated below:
\begin{itemize}
\item no \textit{time-independent} local field in the theory has negative scaling dimension; moreover, the time-independent local field with $\Delta=0=m$ is unique and it is the identity operator. 
\end{itemize}
(We will see that there are several time-dependent fields with $\Delta=0=m$.)

\medskip

Fortunately, the modified assumption does not change any of the results obtained in \cite{Saha:2023hsl} for the mutual OPEs between $S^+_0$, $S^+_1$ and $T$. This statement can be verified by using the restrictions \eqref{7} and \eqref{16}, then repeating the steps to derive those mutual OPEs as elaborated in \cite{Saha:2023hsl} and finally, keeping in mind that the global space-time translation invariance must remain unbroken. Below we collect the mutual OPEs between $S^+_0$ and $S^+_1$ in the $j\epsilon$-form \cite{Saha:2023hsl}:
\begin{align}
&iS_0^+(\mathbf{x})S^+_0(\mathbf{x}_p)\sim0\hspace{5mm};\hspace{5mm} iS_0^+(\mathbf{x})S^+_1(\mathbf{x}_p)\sim\lim\limits_{\epsilon\rightarrow0^+}\text{ }\frac{\bar{z}-\bar{z}_p}{{(\Delta\tilde{z}_p)}}\text{ }S^+_0(\mathbf{x}_p)\nonumber\\
&iS_1^+(\mathbf{x})S^+_0(\mathbf{x}_p)\sim\lim\limits_{\epsilon\rightarrow0^+}\left[\frac{(\bar{z}-\bar{z}_p)^2}{{(\Delta\tilde{z}_p)}}\text{ }\partial_{\bar{z}_p} S^+_0+\frac{\bar{z}-\bar{z}_p}{{(\Delta\tilde{z}_p)}}\text{ }S^+_0\right](\mathbf{x}_p)\\
&i\left\{S_1^+-(t-t_p)S^+_0\right\}(\mathbf{x})S^+_1(\mathbf{x}_p)\sim\lim\limits_{\epsilon\rightarrow0^+}\left[\frac{(\bar{z}-\bar{z}_p)^2}{{(\Delta\tilde{z}_p)}^2}K+\frac{(\bar{z}-\bar{z}_p)^2}{{(\Delta\tilde{z}_p)}}\text{ }\partial_{\bar{z}_p} S^+_1+\frac{\bar{z}-\bar{z}_p}{{(\Delta\tilde{z}_p)}}\text{ }2S^+_1\right](\mathbf{x}_p)\nonumber
\end{align}
where $K$ is a constant not fixed by symmetry.

\medskip

We now proceed to find the mutual OPEs involving the $S^+_2$ field. The general form of the $S^+_0$ OPE \eqref{5}, the modified first assumption, the relation \eqref{9} and the form of the $S^+_0S^+_1$ OPE together completely fix the singular part of the $S^+_0S^+_2$ OPE to be:
\begin{align}
iS_0^+(\mathbf{x})S^+_2(\mathbf{x}_p)\sim\lim\limits_{\epsilon\rightarrow0^+}\text{ }\frac{\bar{z}-\bar{z}_p}{{(\Delta\tilde{z}_p)}}\text{ }S^+_1(\mathbf{x}_p)\label{59}
\end{align}
Using the OPE (bosonic) commutativity property, the $S^+_2S^+_0$ OPE is readily obtained as:
\begin{align}
iS_2^+(\mathbf{x})S^+_0(\mathbf{x}_p)\sim\lim\limits_{\epsilon\rightarrow0^+}&\left[\frac{\left(\bar{z}-\bar{z}_p\right)^3}{{(\Delta\tilde{z}_p)}}\text{ }\frac{1}{2}\bar{\partial}^2_p S^+_1+\frac{\left(\bar{z}-\bar{z}_p\right)^2}{{(\Delta\tilde{z}_p)}}\text{ }\bar{\partial}_p S^+_1+\frac{\bar{z}-\bar{z}_p}{{(\Delta\tilde{z}_p)}}\text{ }S^+_1\right.\nonumber\\
&\left.\hspace{29mm}+(t-t_p)\left\{\frac{(\bar{z}-\bar{z}_p)^2}{{(\Delta\tilde{z}_p)}}\text{ }\bar{\partial}_{p} S^+_0+\frac{\bar{z}-\bar{z}_p}{{(\Delta\tilde{z}_p)}}\text{ }S^+_0\right\}\right](\mathbf{x}_p)
\end{align}
where the OPE gets truncated by virtue of the restrictions \eqref{6}, \eqref{7}, \eqref{15} and \eqref{16}.

\medskip

Similarly, the general form of the $S^+_1$ OPE \eqref{14} and the modified first assumption together with the relation \eqref{9}, the $S^+_1S^+_1$ and the $S^+_0S^+_2$ OPEs determine the singular structure of the $S^+_1S^+_2$ OPE as ($S^+_2$ has $\bar{h}=-\frac{3}{2}$):
\begin{align*}
iS_1^+(\mathbf{x})S^+_2(\mathbf{x}_p)\sim\lim\limits_{\epsilon\rightarrow0^+}\left[\frac{(\bar{z}-\bar{z}_p)^2}{{(\Delta\tilde{z}_p)}^2}Kt+\frac{(\bar{z}-\bar{z}_p)^2}{{(\Delta\tilde{z}_p)}}\text{ }\bar{\partial}_{p} S^+_2+\frac{\bar{z}-\bar{z}_p}{{(\Delta\tilde{z}_p)}}\text{ }3S^+_2+(t-t_p)\frac{\bar{z}-\bar{z}_p}{{(\Delta\tilde{z}_p)}}\text{ }S^+_1\right](\mathbf{x}_p)
\end{align*}
It is now evident that the $\langle S_1^+(\mathbf{x})S^+_2(\mathbf{x}_p)\rangle$ correlator can not be time-translation invariant unless $K=0$ ; so, the final form of the above OPE reduces to:
\begin{align}
iS_1^+(\mathbf{x})S^+_2(\mathbf{x}_p)\sim\lim\limits_{\epsilon\rightarrow0^+}\left[\frac{(\bar{z}-\bar{z}_p)^2}{{(\Delta\tilde{z}_p)}}\text{ }\bar{\partial}_{p} S^+_2+\frac{\bar{z}-\bar{z}_p}{{(\Delta\tilde{z}_p)}}\text{ }3S^+_2+(t-t_p)\frac{\bar{z}-\bar{z}_p}{{(\Delta\tilde{z}_p)}}\text{ }S^+_1\right](\mathbf{x}_p)\label{13}
\end{align}

\medskip

To find the $S^+_2S^+_1$ OPE using the OPE commutativity property from \eqref{13}, we first need to know if the resultant Taylor series gets truncated into a polynomial in $(\bar{z}-\bar{z}_p)$. Recalling \eqref{15}, we observe that the OPE \eqref{59} satisfies:
\begin{align*}
S_0^+\bar{\partial}^{3+N} S^+_2\sim0\hspace{2.5mm}{(N\in\mathbb{N})}\hspace{5mm}\Longrightarrow\hspace{5mm}\left(\bar{\partial}^{3+N} S^+_2\right)S^+_0\sim0
\end{align*}
Now, we consider an arbitrary operator product $\bar{\partial}_1^{3+N} S^+_2(\mathbf{x}_1)S^+_0(\mathbf{x}_2)\Phi(\mathbf{x}_3)$ (where the field $\Phi$ is mutually local with both $S^+_2$ and $S^+_0$) and apply the OPE associativity property\footnote{Following the 2D Euclidean CFT \cite{Zamolodchikov:1989mz}, we assume the associativity property as a consistency condition for the Carrollian conformal OPEs in this work.} in two different ways such that both the resulting series are convergent for the following ordering of the flat Carrollian norms: $\left|\vec{x}_1-\vec{x}_2\right|<\left|\vec{x}_2-\vec{x}_3\right|<\left|\vec{x}_1-\vec{x}_3\right|$. Treating $(z_2-z_3)$ and $(z_1-z_3)$ as the two independent variables, we easily reach the following restriction on $S^+_2$ , analogous to \eqref{6} and \eqref{15}:
\begin{align}
\bar{\partial}^4 S^+_2(t,z,\bar{z})\Phi(\mathbf{x}_p)\sim0\label{19}
\end{align}
implying that an $S^+_2\Phi$ OPE will be at most a cubic order polynomial in $(\bar{z}-\bar{z}_p)$. As a consequence and remembering that we have also postulated that the $S^+_2\Phi$ OPEs have only meromorphic pole singularities while being anti-analytic so that $S^+_2,S^+_1,S^+_0$ are mutually local and that $z,\bar{z}$ are treated independently, inside a correlator the $S^+_{2e}$ part can be decomposed as:
\begin{align}
&S^+_{2e}(t,z,\bar{z})=-k^{(+2)}_e(t,z,\bar{z})+3\bar{z}k^{(+1)}_e(t,z,\bar{z})-3\bar{z}^2k^{(0)}_e(t,z,\bar{z})+\bar{z}^3k^{(-1)}_e(t,z,\bar{z})\label{20}\\
\text{such that}\hspace{2.5mm}&\langle\bar{\partial} k^a_e(t,z,\bar{z})X\rangle=-i\sum\limits_{p=1}^n\text{ }\theta(t-t_p)\text{ }\left[\text{contact terms on $S^2$}\right]\nonumber
\end{align}
just like \eqref{4} and \eqref{18}. The holomorphic weight of all four $k^a_e$ as well as the field $S^+_2$ is $h=\frac{1}{2}$. 

\medskip   
 
With the restriction \eqref{19} in mind, we can immediately write the $S^+_2S^+_1$ OPE from \eqref{13} using the OPE commutativity property as below:
\begin{align}
iS_2^+(\mathbf{x})S^+_1(\mathbf{x}_p)&\sim\lim\limits_{\epsilon\rightarrow0^+}\left[\frac{\left(\bar{z}-\bar{z}_p\right)^3}{{(\Delta\tilde{z}_p)}}\text{ }\frac{1}{2}\bar{\partial}^2_p S^+_2+\frac{\left(\bar{z}-\bar{z}_p\right)^2}{{(\Delta\tilde{z}_p)}}\text{ }2\bar{\partial}_p S^+_2+\frac{\bar{z}-\bar{z}_p}{{(\Delta\tilde{z}_p)}}\text{ }3S^+_2\right.\nonumber\\
&\left.+(t-t_p)\left\{\frac{(\bar{z}-\bar{z}_p)^2}{{(\Delta\tilde{z}_p)}}\text{ }\bar{\partial}_{p} S^+_1+\frac{\bar{z}-\bar{z}_p}{{(\Delta\tilde{z}_p)}}\text{ }2S^+_1\right\}+\frac{1}{2}(t-t_p)^2\text{ }\frac{\bar{z}-\bar{z}_p}{{(\Delta\tilde{z}_p)}}\text{ }S^+_0\right](\mathbf{x}_p)\label{21}
\end{align}
which is also seen to be consistent with the decompositions \eqref{17} and \eqref{20}. 

\medskip

Finally, we proceed to find the $S^+_2S^+_2$ OPE with all the above information at our disposal.

\medskip

\subsection{The $S^+_2S^+_2$ OPE}
We first make the following ansatz directly in the $j\epsilon$-form for the $S^+_2S^+_2$ OPE that is consistent with the decomposition \eqref{17} and the restriction \eqref{19}:
\begin{align}
iS_2^+(\mathbf{x})S^+_2(\mathbf{x}_p)\sim\lim\limits_{\epsilon\rightarrow0^+}&\left[\sum_{r=0}^3\sum_{s\geq1}\frac{(\bar{z}-\bar{z}_p)^r}{{(\Delta\tilde{z}_p)}^s}\text{ }A_{r,s}+(t-t_p)\left\{\frac{(\bar{z}-\bar{z}_p)^2}{{(\Delta\tilde{z}_p)}}\text{ }\bar{\partial}_{p} S^+_2+\frac{\bar{z}-\bar{z}_p}{{(\Delta\tilde{z}_p)}}\text{ }3S^+_2\right\}\right.\nonumber\\
&\left.\hspace{36mm}+\frac{1}{2}(t-t_p)^2\text{ }\frac{\bar{z}-\bar{z}_p}{{(\Delta\tilde{z}_p)}}\text{ }S^+_1\right](\mathbf{x}_p)\label{22}
\end{align}
where $A_{r,s}(\mathbf{x}_p)$ are yet undetermined fields mutually local with $S^+_2,S^+_1,S^+_0$. Now, due to the restriction \eqref{9}, singular part of $S_2^+(\mathbf{x})\partial_{t_p}S^+_2(\mathbf{x}_p)$ obtained from this ansatz must completely match with that of the $S_2^+(\mathbf{x})S^+_1(\mathbf{x}_p)$ OPE given by \eqref{21}. Only the $\mathcal{O}\left((t-t_p)^0\right)$ terms, while comparing, give non-trivial constraints that are listed below (with $\dot{}$ representing time-derivative):
\begin{align}
&\dot{A}_{r,s\geq2}\sim0\hspace{5mm},\hspace{5mm}\dot{A}_{0,1}\sim0\nonumber\\
&\dot{A}_{3,1}\sim\frac{1}{2}\bar{\partial}^2_p S^+_2\hspace{5mm},\hspace{5mm}\dot{A}_{2,1}\sim3\bar{\partial}_p S^+_2\hspace{5mm},\hspace{5mm}\dot{A}_{1,1}\sim6S^+_2\label{23} 
\end{align}
The first line says that all the local fields ${A}_{r,s\geq2}$ and ${A}_{0,1}$ are time-independent (in the OPE limit). Moreover, since the l.h.s. of the OPE \eqref{22} has a total scaling dimension $\Delta=-2$ , all of these fields have negative scaling dimensions. Hence, by the modified first assumption, we set all of them zero.

\medskip

More interestingly, the requirements \eqref{23} reveal that, for the $S^+_2S^+_2$ OPE to be consistent, there must exist a local field $A_{1,1}$ that, in the OPE limit, satisfies $\dot{A}_{1,1}\sim6S^+_2$ which is a condition analogous to \eqref{9}. Let us denote the local field as $S^+_3$ that obeys:
\begin{align}
\left(\partial_t S^+_3-S^+_2\right)(t,z,\bar{z})\Phi(\mathbf{x}_p)\sim0\label{24}
\end{align}
The dimensions of the field $S^+_3$ then are $(\Delta,m)=(-2,2)$.

\medskip

It needs to be emphasized that, unlike $S^+_2$, we did not need to postulate the existence of the local field $S^+_3$. Rather its existence is automatically demanded if the $S^+_2S^+_2$ OPE is to be consistent. This can be interpreted as the non-closure of the mode-algebra of the three fields $S^+_2,S^+_1,S^+_0$ alone.

\medskip

Finally, we write down the $S^+_2S^+_2$ OPE below:
\begin{align}
iS_2^+(\mathbf{x})S^+_2(\mathbf{x}_p)&\sim\lim\limits_{\epsilon\rightarrow0^+}\left[\frac{(\bar{z}-\bar{z}_p)^3}{{(\Delta\tilde{z}_p)}}\text{ }\left(K_1+\frac{1}{2}\bar{\partial}^2_pS^+_3\right)+\frac{(\bar{z}-\bar{z}_p)^2}{{(\Delta\tilde{z}_p)}}\text{ }3\bar{\partial}_pS^+_3+\frac{\bar{z}-\bar{z}_p}{{(\Delta\tilde{z}_p)}}\text{ }6S^+_3\right.\nonumber\\
&\left.+(t-t_p)\left\{\frac{(\bar{z}-\bar{z}_p)^2}{{(\Delta\tilde{z}_p)}}\text{ }\bar{\partial}_{p} S^+_2+\frac{\bar{z}-\bar{z}_p}{{(\Delta\tilde{z}_p)}}\text{ }3S^+_2\right\}+\frac{1}{2}(t-t_p)^2\text{ }\frac{\bar{z}-\bar{z}_p}{{(\Delta\tilde{z}_p)}}\text{ }S^+_1\right](\mathbf{x}_p)\label{26}
\end{align}
with $K_1$ being a constant not yet fixed. It is clear from this OPE that the commutators of only the modes contained in $S^+_{2e}$ will involve new modes appearing in $S^{+}_{3e}$ that is defined below analogously to \eqref{17} and consistently with \eqref{24}: 
\begin{align}
&S^+_3(t,z,\bar{z})=S^+_{3e}(t,z,\bar{z})+tS^+_{2e}(t,z,\bar{z})+\frac{t^2}{2}S^+_{1e}(t,z,\bar{z})+\frac{t^3}{3!}S^+_0(t,z,\bar{z})\\
\Longrightarrow\hspace{5mm}&{\partial_t}\langle S^+_{3e}(t,z,\bar{z})X\rangle=\left[\text{temporal contact terms}\right]\nonumber
\end{align}

\medskip

We shall now investigate on the properties of the local field $S^+_3$ and discover that a tower of local fields $S^+_k$ ($k\geq4$) will be required to automatically exist for the consistency of the OPEs.

\medskip

\section{The Tower of Fields $S^+_k$}\label{s4}
We shall begin by looking into the mutual OPEs of the four local fields $S^+_k$ with $0\leq k\leq 3$ and observe that, for the OPE $S^+_2S^+_3$ to be consistent, a new local field $S^+_4$ must exist. Similarly, for the consistency of the $S^+_2S^+_4$ OPE, another new field $S^+_5$ is required to automatically exist. As is anticipated, this sort of argument will recursively generate the whole tower of the fields $S^+_k$ ($k\geq4$).

\medskip

\subsection{The field $S^+_3$}
Following the steps taken (and remembering the modified first assumption) in the previous section to construct the OPEs involving $S^+_2$, we directly write down the following OPEs for $S^+_3$ (it has $\bar{h}=-2$):
\begin{align}
&iS_0^+(\mathbf{x})S^+_3(\mathbf{x}_p)\sim\lim\limits_{\epsilon\rightarrow0^+}\text{ }\frac{\bar{z}-\bar{z}_p}{{(\Delta\tilde{z}_p)}}\text{ }S^+_2(\mathbf{x}_p)\\
&iS_1^+(\mathbf{x})S^+_3(\mathbf{x}_p)\sim\lim\limits_{\epsilon\rightarrow0^+}\left[\frac{(\bar{z}-\bar{z}_p)^2}{{(\Delta\tilde{z}_p)}}\text{ }\bar{\partial}_{p} S^+_3+\frac{\bar{z}-\bar{z}_p}{{(\Delta\tilde{z}_p)}}\text{ }4S^+_3+(t-t_p)\frac{\bar{z}-\bar{z}_p}{{(\Delta\tilde{z}_p)}}\text{ }S^+_2\right](\mathbf{x}_p)
\end{align}
Together with the restriction \eqref{19}, the first OPE implies, similarly as derived for $S^+_2$, that:
\begin{align}
\bar{\partial}^5 S^+_3(t,z,\bar{z})\Phi(\mathbf{x}_p)\sim0\label{25}
\end{align}
which means that an $S^+_3\Phi$ OPE will be at most a quartic polynomial in $(\bar{z}-\bar{z}_p)$. Since, the singularity structure of such an OPE must be the same as that of the $S^+_0\Phi$ and $S^+_1\Phi$ OPEs, the condition \eqref{25} allows us to decompose $S^+_{3e}$ into five objects $l^a_e(t,z,\bar{z})$ whose correlators are both time-independent and holomorphic in the OPE limit, analogous to \eqref{20}.

\medskip

Now, to find the $S^+_2S^+_3$ OPE, let us first construct the following ansatz consistent with the conditions \eqref{17} and \eqref{19}:
\begin{align*}
iS_2^+(\mathbf{x})S^+_3(\mathbf{x}_p)\sim\lim\limits_{\epsilon\rightarrow0^+}&\left[\sum_{r=0}^3\sum_{s\geq1}\frac{(\bar{z}-\bar{z}_p)^r}{{(\Delta\tilde{z}_p)}^s}\text{ }B_{r,s}+(t-t_p)\left\{\frac{(\bar{z}-\bar{z}_p)^2}{{(\Delta\tilde{z}_p)}}\text{ }\bar{\partial}_{p} S^+_3+\frac{\bar{z}-\bar{z}_p}{{(\Delta\tilde{z}_p)}}\text{ }4S^+_3\right\}\right.\nonumber\\
&\left.\hspace{36mm}+\frac{1}{2}(t-t_p)^2\text{ }\frac{\bar{z}-\bar{z}_p}{{(\Delta\tilde{z}_p)}}\text{ }S^+_2\right](\mathbf{x}_p)
\end{align*}
where $B_{r,s}(\mathbf{x}_p)$ are yet undetermined local fields. Now, obeying the restriction \eqref{24}, we match the singular part of $S_2^+(\mathbf{x})\partial_{t_p}S^+_3(\mathbf{x}_p)$ obtained from this ansatz with that of the $S_2^+(\mathbf{x})S^+_2(\mathbf{x}_p)$ OPE given by \eqref{26} to find that, again, only the $\mathcal{O}\left((t-t_p)^0\right)$ terms give non-trivial constraints collected below:
\begin{align}
&\dot{B}_{r,s\geq2}\sim0\hspace{5mm},\hspace{5mm}\dot{B}_{0,1}\sim0\nonumber\\
&\dot{B}_{3,1}\sim\frac{1}{2}\bar{\partial}^2_p S^+_3+K_1\hspace{5mm},\hspace{5mm}\dot{B}_{2,1}\sim4\bar{\partial}_p S^+_3\hspace{5mm},\hspace{5mm}\dot{B}_{1,1}\sim10S^+_3\label{27} 
\end{align}
Clearly, all of the time-independent local fields ${B}_{r,s\geq2}$ and ${B}_{0,1}$ have negative scaling dimensions; hence, all of them are set zero by the modified first assumption. But, the conditions \eqref{27} demand that there must exist a local field $S^+_4$ such that:
\begin{align}
\left(\partial_t S^+_4-S^+_3\right)(t,z,\bar{z})\Phi(\mathbf{x}_p)\sim0\label{58}
\end{align}
The dimensions of $S^+_4$ hence must be $(\Delta,m)=(-3,2)$. 

\medskip

The $S^+_2S^+_3$ OPE then finally is:
\begin{align}
iS_2^+(\mathbf{x})S^+_3(\mathbf{x}_p)&\sim\lim\limits_{\epsilon\rightarrow0^+}\left[\frac{(\bar{z}-\bar{z}_p)^3}{{(\Delta\tilde{z}_p)}}\text{ }\frac{1}{2}\bar{\partial}^2_pS^+_4+\frac{(\bar{z}-\bar{z}_p)^2}{{(\Delta\tilde{z}_p)}}\text{ }4\bar{\partial}_pS^+_4+\frac{\bar{z}-\bar{z}_p}{{(\Delta\tilde{z}_p)}}\text{ }10S^+_4\right.\nonumber\\
&\left.+(t-t_p)\left\{\frac{(\bar{z}-\bar{z}_p)^2}{{(\Delta\tilde{z}_p)}}\text{ }\bar{\partial}_{p} S^+_3+\frac{\bar{z}-\bar{z}_p}{{(\Delta\tilde{z}_p)}}\text{ }4S^+_3\right\}+\frac{1}{2}(t-t_p)^2\text{ }\frac{\bar{z}-\bar{z}_p}{{(\Delta\tilde{z}_p)}}\text{ }S^+_2\right](\mathbf{x}_p)
\end{align}
with $K_1=0$ also, to keep the time-invariance of the $\langle S_2^+(\mathbf{x})S^+_3(\mathbf{x}_p)\rangle$ correlator intact.

\medskip

It is by now apparent that this procedure will recursively reveal the automatic existence of a tower of local fields $S^+_{k+1}$ ($k\geq2$) just from the requirement of the consistency of the $S^+_2S^+_k$ OPE such that:
\begin{align}
\left(\partial_t S^+_{k+1}-S^+_k\right)(t,z,\bar{z})\Phi(\mathbf{x}_p)\sim0\label{28}
\end{align}
if we merely postulate that the $1+2$D CarrCFT contains the field $S^+_2$ in the first place.

\medskip

We shall now provide a recursive construction of the general $S^+_2S^+_k$ OPE.

\medskip

\subsection{The fields $S^+_k$}
The relation \eqref{28} between the fields $S^+_k$ and $S^+_{k+1}$ leads to:
\begin{align*}
\left(\partial^{k}_t S^+_{k}-S^+_0\right)(t,z,\bar{z})\Phi(\mathbf{x}_p)\sim0
\end{align*}
that implies that the field $S^+_k$ must have dimensions $(\Delta,m)=(1-k,2)$. So, the following OPEs are immediately constructed:
\begin{align}
&iS_0^+(\mathbf{x})S^+_{k+1}(\mathbf{x}_p)\sim\lim\limits_{\epsilon\rightarrow0^+}\text{ }\frac{\bar{z}-\bar{z}_p}{{(\Delta\tilde{z}_p)}}\text{ }S^+_k(\mathbf{x}_p)\label{30}\\
&iS_1^+(\mathbf{x})S^+_{k+1}(\mathbf{x}_p)\sim\lim\limits_{\epsilon\rightarrow0^+}\left[\frac{(\bar{z}-\bar{z}_p)^2}{{(\Delta\tilde{z}_p)}}\text{ }\bar{\partial}_{p} S^+_{k+1}+\frac{\bar{z}-\bar{z}_p}{{(\Delta\tilde{z}_p)}}\text{ }(k+2)S^+_{k+1}+(t-t_p)\frac{\bar{z}-\bar{z}_p}{{(\Delta\tilde{z}_p)}}\text{ }S^+_k\right](\mathbf{x}_p)\label{31}
\end{align}
As can be deduced by induction from the knowledge of \eqref{19}, the first OPE implies that:
\begin{align}
\bar{\partial}^{k+2} S^+_k(t,z,\bar{z})\Phi(\mathbf{x}_p)\sim0\label{29}
\end{align}
which translates into the fact that an $S^+_k\Phi$ OPE will be at most a $(k+1)$-th order polynomial in $(\bar{z}-\bar{z}_p)$.

\medskip

The relation \eqref{28} allows for the following decomposition of $S^+_k$ inside a correlator:
\begin{align}
&S^+_k(t,z,\bar{z})=\sum_{r=1}^k \frac{t^{k-r}}{(k-r)!}\text{ } S^+_{r(e)}(t,z,\bar{z})+\frac{t^k}{k!}S^+_0(t,z,\bar{z})\label{33}\\
\Longrightarrow\hspace{5mm}&{\partial_t}\langle S^+_{r(e)}(t,z,\bar{z})X\rangle=\left[\text{temporal contact terms}\right]\nonumber
\end{align}
while the condition \eqref{29} permits further decomposition of $S^+_{k(e)}$ inside a correlator as shown below:
\begin{align}
&S^+_{k(e)}(t,z,\bar{z})=\frac{1}{(k+1)!}\sum^{k+1}_{s=0}{k+1 \choose s}(-)^{k+1-s}\bar{z}^{s}\text{ } H^k_{\frac{k+1}{2}-s}(t,z,\bar{z})\label{39}\\
&S^+_0(t,z,\bar{z})=\bar{z}H^0_{-\frac{1}{2}}(t,z,\bar{z})-H^0_{\frac{1}{2}}(t,z,\bar{z})\nonumber\\
\text{such that}\hspace{2.5mm}&\langle\bar{\partial} H^k_{{\frac{k+1}{2}-s}}(t,z,\bar{z})X\rangle=-i\sum\limits_{p=1}^n\text{ }\theta(t-t_p)\text{ }\left[\text{contact terms on $S^2$}\right]\nonumber
\end{align} 
Such a decomposition is possible because of the holomorphic singularity structure and $z,\bar{z}$ being independent. As before, $S^+_{k(e)}$ are local Carrollian fields but are merely collections of modes. The holomorphic weight of $H^k_{{\frac{k+1}{2}-s}}$ as well as the field $S^+_k$ is $h=\frac{3-k}{2}$.

\medskip

Now, we make an ansatz for the $S^+_2S^+_k$ OPE following the same steps as before but omitting the terms that will be eventually set zero by the modified first assumption:
\begin{align*}
iS_2^+(\mathbf{x})&S^+_k(\mathbf{x}_p)\sim\lim\limits_{\epsilon\rightarrow0^+}\left[\frac{(\bar{z}-\bar{z}_p)^3}{{(\Delta\tilde{z}_p)}}\text{ }A^{(k)}_3+\frac{(\bar{z}-\bar{z}_p)^2}{{(\Delta\tilde{z}_p)}}\text{ }A^{(k)}_2+\frac{\bar{z}-\bar{z}_p}{{(\Delta\tilde{z}_p)}}\text{ }A^{(k)}_1\right.\nonumber\\
&\left.+(t-t_p)\left\{\frac{(\bar{z}-\bar{z}_p)^2}{{(\Delta\tilde{z}_p)}}\text{ }\bar{\partial}_{p} S^+_k+\frac{\bar{z}-\bar{z}_p}{{(\Delta\tilde{z}_p)}}\text{ }(k+1)S^+_k\right\}+\frac{1}{2}(t-t_p)^2\text{ }\frac{\bar{z}-\bar{z}_p}{{(\Delta\tilde{z}_p)}}\text{ }S^+_{k-1}\right](\mathbf{x}_p)
\end{align*}
with $A^{(k)}_p$ being some local fields. The $S_2^+(\mathbf{x})\dot{S}^+_k(\mathbf{x}_p)$ OPE obtained from this ansatz must be the same as the ansatz for the $S_2^+(\mathbf{x})S^+_{k-1}(\mathbf{x}_p)$ OPE due to the relation \eqref{28}. Again, the non-trivial constraints come only from the $\mathcal{O}\left((t-t_p)^0\right)$ terms. These give rise to the following recursive system for the local fields $A^{(k)}_p$:
\begin{align*}
&\dot{A}^{(k)}_3\sim A^{(k-1)}_3\hspace{5mm},\hspace{5mm}\dot{A}^{(k)}_2-\bar{\partial}_{p} S^+_k\sim A^{(k-1)}_2\hspace{5mm},\hspace{5mm}\dot{A}^{(k)}_1-(k+1)S^+_k\sim A^{(k-1)}_1\\
\text{with seeds}\hspace{5mm}&{A}^{(2)}_3=\frac{1}{2}\bar{\partial}^2S^+_3\hspace{5mm},\hspace{5mm}{A}^{(2)}_2=3\bar{\partial}S^+_3\hspace{5mm},\hspace{5mm}{A}^{(2)}_1=6S^+_3
\end{align*}

\medskip

We demonstrate the solution of the $A_1$ recursive system below:
\begin{align*}
\dot{A}^{(k)}_1\sim A^{(k-1)}_1+(k+1)S^+_k\hspace{2.5mm}\Rightarrow\hspace{2.5mm}&\partial^2_t {A}^{(k)}_1\sim A^{(k-2)}_1+(k+1)S^+_{k-1}+kS^+_{k-1}\\
\Rightarrow\hspace{2.5mm}&\partial_t^{k-2}{A}^{(k)}_1\sim A^{(2)}_1+[(k+1)+k+(k-1)+\ldots+4]S^+_{3}\\
\Rightarrow\hspace{2.5mm}&\partial_t^{k-2}{A}^{(k)}_1\sim \frac{1}{2}(k+1)(k+2)S^+_{3}\\
\Rightarrow\hspace{2.5mm}&{A}^{(k)}_1\sim \frac{1}{2}(k+1)(k+2)S^+_{k+1}
\end{align*}
The inversion of the time-derivatives are unique because of the severe restrictions imposed by the modified first assumption. This can also be motivated from the specific examples studied above.

\medskip

The unique solutions to all the three recursions are similarly obtained to be:
\begin{align*}
{A}^{(k)}_3\sim\frac{1}{2}\bar{\partial}^2S^+_{k+1}\hspace{5mm},\hspace{5mm}{A}^{(k)}_2\sim(k+1)\bar{\partial}S^+_{k+1}\hspace{5mm},\hspace{5mm}{A}^{(k)}_1\sim\frac{1}{2}(k+1)(k+2)S^+_{k+1}
\end{align*}
consistent with all the previously considered specific cases. The $S^+_2S^+_k$ OPE is finally noted below:
\begin{align}
iS_2^+(\mathbf{x})&S^+_k(\mathbf{x}_p)\sim\lim\limits_{\epsilon\rightarrow0^+}\left[\frac{(\bar{z}-\bar{z}_p)^3}{{(\Delta\tilde{z}_p)}}\text{ }\frac{1}{2}\bar{\partial}_p^2S^+_{k+1}+\frac{(\bar{z}-\bar{z}_p)^2}{{(\Delta\tilde{z}_p)}}\text{ }(k+1)\bar{\partial}_pS^+_{k+1}+\frac{\bar{z}-\bar{z}_p}{{(\Delta\tilde{z}_p)}}\text{ }\frac{1}{2}(k+1)(k+2)S^+_{k+1}\right.\nonumber\\
&\left.+(t-t_p)\left\{\frac{(\bar{z}-\bar{z}_p)^2}{{(\Delta\tilde{z}_p)}}\text{ }\bar{\partial}_{p} S^+_k+\frac{\bar{z}-\bar{z}_p}{{(\Delta\tilde{z}_p)}}\text{ }(k+1)S^+_k\right\}+\frac{1}{2}(t-t_p)^2\text{ }\frac{\bar{z}-\bar{z}_p}{{(\Delta\tilde{z}_p)}}\text{ }S^+_{k-1}\right](\mathbf{x}_p)\label{32}
\end{align}

\medskip

With the above derivation as a warm-up, we finally attempt to find the most general $S^+_kS^+_l$ OPE.

\medskip

\subsection{The general $S^+_kS^+_l$ OPE}\label{s4.3}
We shall find the general $S^+_kS^+_l$ OPE via a recursive method analogous to the one demonstrated above. To find the seeds of the recursions, we first note down the three cases with $l=0,1,2$. This can be achieved using the OPE commutativity property from the known OPEs $S^+_lS^+_k$ with $l=0,1,2$, i.e. from the OPEs \eqref{30}, \eqref{31} and \eqref{32} respectively. While employing the OPE commutativity property, we need to recall that an $S^+_k\Phi$ OPE is a $(k+1)$-th order polynomial in $(\bar{z}-\bar{z}_p)$ and a $k$-th order polynomial in $(t-t_p)$, due to respectively \eqref{29} and \eqref{33}. Also, since in an $S^+_kS^+_l$ OPE only the $\mathcal{O}\left((t-t_p)^0\right)$ terms are new in the sense that the $\mathcal{O}\left((t-t_p)^r\right)$ (with $r\in\mathbb{N}$) terms' coefficients have already been the $\mathcal{O}\left((t-t_p)^0\right)$ term in the $S^+_{k-r}S^+_l$ OPE, we shall only explicitly write the $\mathcal{O}\left((t-t_p)^0\right)$ term from now on in a general OPE. The results are (with h.o.t. denoting the $\mathcal{O}\left((t-t_p)^r\right)$ terms with $r\geq1$):
\begin{align}
&iS_k^+(\mathbf{x})S^+_0(\mathbf{x}_p)\sim\lim\limits_{\epsilon\rightarrow0^+}\text{ }\sum_{m=0}^k\frac{\left(\bar{z}-\bar{z}_p\right)^{m+1}}{{(\Delta\tilde{z}_p)}}\text{ }\frac{1}{m!}\bar{\partial}^m_p S^+_{k-1}(\mathbf{x}_p)+\text{(h.o.t.)}\label{35}\\
&iS_k^+(\mathbf{x})S^+_{1}(\mathbf{x}_p)\sim\lim\limits_{\epsilon\rightarrow0^+}\text{ }\sum_{m=0}^k\frac{\left(\bar{z}-\bar{z}_p\right)^{m+1}}{{(\Delta\tilde{z}_p)}}\text{ }\frac{k+1-m}{m!}\text{ }\bar{\partial}^m_p S^+_{k}(\mathbf{x}_p)+\text{(h.o.t.)}\\
&iS_k^+(\mathbf{x})S^+_{2}(\mathbf{x}_p)\sim\lim\limits_{\epsilon\rightarrow0^+}\text{ }\sum_{m=0}^k\frac{\left(\bar{z}-\bar{z}_p\right)^{m+1}}{{(\Delta\tilde{z}_p)}}\text{ }\frac{(k+1-m)(k+2-m)}{2\cdot m!}\text{ }\bar{\partial}^m_p S^+_{k+1}(\mathbf{x}_p)+\text{(h.o.t.)}\label{34}
\end{align}

\medskip

With this knowledge, we shall now find the $S^+_kS^+_3$ OPE in exactly the same way we derived the $S^+_2S^+_3$ OPE exploiting the $S^+_2S^+_2$ OPE. Let the ansatz for the $S^+_kS^+_3$ OPE be:
\begin{align*}
&iS_k^+(\mathbf{x})S^+_{3}(\mathbf{x}_p)\sim\lim\limits_{\epsilon\rightarrow0^+}\left[\sum_{m=0}^k\frac{\left(\bar{z}-\bar{z}_p\right)^{m+1}}{{(\Delta\tilde{z}_p)}}\text{ }B^{(k)}_m+(t-t_p)\sum_{m=0}^{k-1}\frac{\left(\bar{z}-\bar{z}_p\right)^{m+1}}{{(\Delta\tilde{z}_p)}}\text{ }B^{(k-1)}_m\right](\mathbf{x}_p)+\text{(h.o.t.)}^\prime
\end{align*} 
where $B^{(r)}_m(\mathbf{x}_p)$ are the local fields to be determined via recursion. Now, due to the relation \eqref{24}, the $iS_k^+(\mathbf{x})\dot{S}^+_{3}(\mathbf{x}_p)$ OPE derived from this ansatz must be the same as the OPE \eqref{34}. This gives rise to the following recursion relation:
\begin{align*}
&\dot{B}^{(k)}_m\sim B^{(k-1)}_m\frac{(k+1-m)(k+2-m)}{2\cdot m!}\text{ }\bar{\partial}^m_p S^+_{k+1}\\
\Rightarrow\hspace{2.5mm}&\partial^2_t{B}^{(k)}_m\sim B^{(k-2)}_m+\frac{(k+1-m)(k+2-m)}{2\cdot m!}\text{ }\bar{\partial}^m_p S^+_{k}+\frac{(k-m)(k+1-m)}{2\cdot m!}\text{ }\bar{\partial}^m_p S^+_{k}\\
\Rightarrow\hspace{2.5mm}&\partial^{k-m+1}_t{B}^{(k)}_m\sim B^{(m-1)}_m+\frac{1}{2\cdot m!}\left[\sum_{n=1}^{k+1-m}n(n+1)\right]\bar{\partial}^m_p S^+_{m+1}
\end{align*}
To solve this recursion, we need a seed. Since we are comparing the coefficients of the $(t-t_p)^0\frac{\left(\bar{z}-\bar{z}_p\right)^{m+1}}{{(\Delta\tilde{z}_p)}}$ terms, we need to remember that such terms only occurs in the $iS_r^+{S}^+_{3}$ OPEs with $r\geq m$. This means that $B^{(m-1)}_m=0$ and that is the seed of the recursion. Thus, we have:
\begin{align*}
&\partial^{k-m+1}_t{B}^{(k)}_m\sim \frac{(k+1-m)(k+2-m)(k+3-m)}{6\cdot m!}\bar{\partial}^m_p S^+_{m+1}\\
\Rightarrow\hspace{2.5mm}&{B}^{(k)}_m\sim \frac{(k+1-m)(k+2-m)(k+3-m)}{6\cdot m!}\text{ }\bar{\partial}^m_p S^+_{k+2}
\end{align*}
So, the $S^+_kS^+_3$ OPE is given by:
\begin{align}
iS_k^+(\mathbf{x})S^+_{3}(\mathbf{x}_p)\sim\lim\limits_{\epsilon\rightarrow0^+}\text{ }\sum_{m=0}^k\frac{\left(\bar{z}-\bar{z}_p\right)^{m+1}}{{(\Delta\tilde{z}_p)}}\text{ }\frac{(k+1-m)_3}{6\cdot m!}\text{ }\bar{\partial}^m_p S^+_{k+2}(\mathbf{x}_p)+\text{(h.o.t.)}
\end{align}
where $(\ldots)_3$ is an upward Pochhammer symbol.

\medskip

In the exactly similar way, we find the $S^+_kS^+_4$ OPE from the knowledge of the $S^+_kS^+_3$ OPE to be:
\begin{align}
iS_k^+(\mathbf{x})S^+_{4}(\mathbf{x}_p)\sim\lim\limits_{\epsilon\rightarrow0^+}\text{ }\sum_{m=0}^k\frac{\left(\bar{z}-\bar{z}_p\right)^{m+1}}{{(\Delta\tilde{z}_p)}}\text{ }\frac{(k+1-m)_4}{24\cdot m!}\text{ }\bar{\partial}^m_p S^+_{k+3}(\mathbf{x}_p)+\text{(h.o.t.)}
\end{align}
and the $S^+_kS^+_5$ OPE from the $S^+_kS^+_4$ OPE as:
\begin{align}
iS_k^+(\mathbf{x})S^+_{5}(\mathbf{x}_p)\sim\lim\limits_{\epsilon\rightarrow0^+}\text{ }\sum_{m=0}^k\frac{\left(\bar{z}-\bar{z}_p\right)^{m+1}}{{(\Delta\tilde{z}_p)}}\text{ }\frac{(k+1-m)_5}{120\cdot m!}\text{ }\bar{\partial}^m_p S^+_{k+4}(\mathbf{x}_p)+\text{(h.o.t.)}\label{36}
\end{align}

\medskip

The form of the explicit examples \eqref{35}-\eqref{36} inspires the following obvious guess for the general OPE $S^+_kS^+_l$:
\begin{align}
iS_k^+(\mathbf{x})S^+_{l}(\mathbf{x}_p)\sim\lim\limits_{\epsilon\rightarrow0^+}\text{ }\sum_{m=0}^k\frac{\left(\bar{z}-\bar{z}_p\right)^{m+1}}{{(\Delta\tilde{z}_p)}}\text{ }\frac{(k+1-m)_l}{l!\cdot m!}\text{ }\bar{\partial}^m_p S^+_{k+l-1}(\mathbf{x}_p)+\text{(h.o.t.)}
\end{align}
That this is indeed true can be quickly verified by noticing that the OPE $S_k^+\dot{S}^+_{l}$ is exactly the same as the OPE $S_k^+{S}^+_{l-1}$, as must hold because of the relation \eqref{28}.

\medskip

For the sake of completeness, the full form of the above OPE, consistent with the decomposition \eqref{33}, is written below:
\begin{align}
iS_k^+(\mathbf{x})S^+_{l}(\mathbf{x}_p)\sim\lim\limits_{\epsilon\rightarrow0^+}\sum_{r=0}^k
\frac{\left(t-t_p\right)^{k-r}}{(k-r)!}\sum_{m=0}^r\frac{\left(\bar{z}-\bar{z}_p\right)^{m+1}}{{(\Delta\tilde{z}_p)}}\text{ }\frac{(r+1-m)_l}{l!\cdot m!}\text{ }\bar{\partial}^m_p S^+_{r+l-1}(\mathbf{x}_p)\label{37}
\end{align}

\medskip

Having derived the general $S^+_kS^+_l$ OPE, we now take on the final goal of this paper: uncovering the symmetry algebra manifest in the mutual OPEs of the tower of fields $\left\{S^+_k\right\}$.  

\medskip

\section{The Symmetry Algebra}\label{s5}
To find the quantum symmetry algebra from the OPEs, one needs to first identify the contributions of the modes to the OPEs. As is implied by the decomposition \eqref{33}, the new modes in a field $S^+_k$ that do not appear in the fields $S^+_{k-r}$ with $1\leq r\leq k$ are all contained in the part $S^+_{k(e)}$. In this sense, the part $S^+_{k(e)}$ is the unique signature of the field $S^+_k$. Similarly, the `new' information content of the $S^+_kS^+_l$ OPE is described by the $S^+_{k(e)}S^+_{l(e)}$ part. 

\medskip

We recall that the part $S^+_{k(e)}$ is not a Carrollian conformal field. So, strictly speaking, the $S^+_{k(e)}S^+_{l(e)}$ part is not a Carrollian conformal OPE. Nevertheless, we shall call it an `OPE' from now on. It can be readily extracted from the $S^+_kS^+_l$ OPE \eqref{37} by comparing the $\mathcal{O}(t^0t_p^0)$ terms from both sides to be:
\begin{align}
iS_{k(e)}^+(t,z,\bar{z})S^+_{l(e)}(t_p,z_p,\bar{z}_p)\sim\lim\limits_{\epsilon\rightarrow0^+}\text{ }\sum_{m=0}^k\frac{\left(\bar{z}-\bar{z}_p\right)^{m+1}}{{(\Delta\tilde{z}_p)}}\text{ }\frac{(k+1-m)_l}{l!\cdot m!}\text{ }\bar{\partial}^m_p S^+_{k+l-1(e)}(t_p,z_p,\bar{z}_p)\label{38}
\end{align}

\medskip

In the explicit $t\rightarrow\infty$ limit where $\frac{1}{\Delta\tilde{z}_p}\equiv \frac{1}{z-{z}_p}$ (and then recalling that any $S^+_{r(e)}$ is time-independent in the OPE limit), this `OPE' is exactly the same as the OPE of the positive-helicity conformally soft gravitons at the tree level of the Einstein gravity in the bulk AFS \cite{Guevara:2021abz}. There, it was obtained by taking the conformally soft limit from the general OPE of two Celestial conformal primary gravitons derived in \cite{Pate:2019lpp} that needed some crucial inputs from the explicit bulk gravity theory. Here, we have reached the conformally soft graviton OPE directly by instead exploiting only the constraints imposed by the Carrollian conformal symmetry along with the two assumptions stated in section \ref{s3.1}. No hint from the underlying bulk theory was needed in our field theory analysis. The Carrollian time-coordinate $t$ played the central role in our construction.

\medskip

Thus, the conformally soft graviton operator $H^k(z,\bar{z})$ of \cite{Guevara:2021abz} that is a primary field in 2D Celestial CFT is the two-dimensional object $iS_{1-k(e)}^+(\infty,z,\bar{z})$ appearing in the Carrollian conformal field $iS_{1-k}^+(t\rightarrow\infty,z,\bar{z})$.

\medskip

Moving back to our goal of finding the symmetry algebra, we now substitute the anti-holomorphic mode-expansion \eqref{39} into the `OPE' \eqref{38} and match the powers of both $\bar{z}$ and $\bar{z}_p$ from both hand sides to find an `OPE' $H^k_{\frac{k+1}{2}-r}(z)H^l_{\frac{l+1}{2}-s}(z_p)$. The dependence on the time- and the anti-holomorphic coordinates are removed since, by \eqref{33} and \eqref{39}, the Carrollian `modes' $H^k_{s^\prime}(t,z,\bar{z})$ are independent of respectively $t$ and $\bar{z}$ in the OPE limit, thus making them effectively holomorphic inside an OPE. While it is straightforward to match the powers of $\bar{z}$, to do the same with $\bar{z}_p$ we must first insert the anti-holomorphic mode-expansion for $\bar{\partial}^m_pS^+_{k+l-1(e)}(z_p,\bar{z}_p)$ on the r.h.s.. Doing these, one finds the following `OPE' (with $\Delta\tilde{z}_p=z-z_p-j\epsilon0^+$ from now on):
\begin{align*}
&-i\frac{{k+1\choose r}}{(k+1)!}\frac{{l+1\choose s}}{(l+1)!} H^k_{\frac{k+1}{2}-r}(z)H^l_{\frac{l+1}{2}-s}(z_p)\\
\sim&\lim\limits_{\epsilon\rightarrow0^+}\text{ }\sum_{m=\text{max}\{0,r-1\}}^k\frac{(-1)^{m+1-r}}{{(\Delta\tilde{z}_p)}}{m+1\choose r}{s+r-1\choose m}\frac{(k+1-m)_l}{l!\cdot m!}\frac{{k+l\choose r+s-1}}{(k+l)!} H^{k+l-1}_{\frac{k+l}{2}-r-s+1}(z_p)
\end{align*}
The summation over $m$ at the r.h.s. of this `OPE' can be done as below (for $r\geq 1$):
\begin{align*}
&\sum_{m=r-1}^k{(-1)^{m+1-r}}{m+1\choose r}{s+r-1\choose m}\frac{(k+1-m)_l}{l!\cdot m!}\\
=\text{ }&(-)^{k+1-r}\frac{(s+r-1)!}{s!r!}\sum_{m=0}^{k+1-r}(m+r){s\choose m}\frac{(l+1)_{k+1-r-m}}{(k+1-r-m)!}(-)^{k+1-r-m}\\
=\text{ }&(-)^{k+1-r}\frac{(s+r-1)!}{s!r!}\times\left[\text{Coeff. of $x^{k+1-r}$ in $[r+x(r+s)](1+x)^{s-l-2}$ }\right]\\
=\text{ }&\frac{(s+r-1)!}{s!r!}\frac{(k+l-s-r+1)!}{(k+1-r)!(l+1-s)!}\text{ }[r(l+1)-s(k+1)]
\end{align*}
that also holds true for $r=0$. 

\medskip

Thus, the above `OPE' reduces to the following simple form:
\begin{align}
iH^k_{a}(z)H^l_{b}(z_p)\sim\lim\limits_{\epsilon\rightarrow0^+}\text{ }[a(l+1)-b(k+1)]\text{ }\frac{H^{k+l-1}_{a+b}(z_p)}{z-z_p-j\epsilon0^+}\label{40}
\end{align}
with $-\frac{k+1}{2}\leq a\leq\frac{k+1}{2}$ and $-\frac{l+1}{2}\leq b\leq\frac{l+1}{2}$ now ($2a,2b\in\mathbb{Z}$). This has precisely the same appearance as a Kac-Moody current OPE in a holomorphic 2D CFT (after explicitly putting $\epsilon\rightarrow0^+$). But the infinite-dimensional Lie algebra underlying such a Kac-Moody current algebra is not standard in the sense that it has no name! Fortunately, as we see below, this Lie algebra can be recast into a standard form.

\medskip

For this purpose, we relabel the `modes' $H^k_{a}$ as:
\begin{align}
w^p_a\equiv \frac{i}{2} H^{2p-3}_a \hspace{5mm}\Longrightarrow\hspace{5mm} p\in\frac{\mathbb{N}}{2}+1\hspace{2.5mm} \text{and}\hspace{2.5mm} 1-p\leq a\leq p-1\label{41}
\end{align}
and get the following `OPE' for the modes $w^p_a$ from \eqref{40}:
\begin{align}
w^p_{a}(z)w^q_{b}(z_p)\sim\lim\limits_{\epsilon\rightarrow0^+}\text{ }[a(q-1)-b(p-1)]\text{ }\frac{w^{p+q-2}_{a+b}(z_p)}{z-z_p-j\epsilon0^+}
\end{align}
that resembles a 2D CFT Kac-Moody current OPE with the Lie algebra being (a sub-algebra of) the $w_{1+\infty}$ algebra \cite{Bakas:1989xu}. The underlying Lie algebra would actually be the `wedge sub-algebra' defined by the restriction \eqref{41} on $a$ \cite{Pope:1991ig} of the full $w_{1+\infty}$ algebra with unconstrained $a\in\mathbb{Z}$ \cite{Bakas:1989xu}. 

\medskip

In the context of 2D Celestial CFT, the algebra of conformally soft gravitons was re-expressed as this same Kac-Moody algebra in \cite{Strominger:2021lvk}. In this work, the central term $w^1_0$ of \cite{Strominger:2021lvk} is set zero to respect the time-translation invariance. But more importantly, we do not interpret the mode-redefinition \eqref{41} as a (discrete) light-transformation as opposed to the descriptions presented in \cite{Strominger:2021lvk,Himwich:2021dau}.

\medskip

We shall now explicitly show that the `OPE' \eqref{41} actually gives rise to the above said current-algebra symmetry in the CarrCFT (rather than the suggestive resemblance) as the algebra of the modes. This check is important because the CarrCFT technology has some differences with those in the usual 2D CFT. In particular, the mode-commutation relation is shown in \cite{Saha:2023hsl,Saha:2022gjw} to be related with the corresponding OPE via a complex-contour integral in CarrCFT without the need to perform any radial-quantization. It is the temporal step-function- or the $j\epsilon$- prescriptions of the CarrCFT OPEs instead that play the crucial role in establishing such a relation. 

\medskip

First we notice that the objects $H^k_{a}(z)$ can be holomorphic mode-expanded inside a correlator as below, by remembering that its holomorphic weight is $h=\frac{3-k}{2}$:
\begin{align}
H^k_a(z)=\sum_{n\in\mathbb{Z}+\{\frac{k+1}{2}\}}H^k_{a;n} z^{-n-\frac{3-k}{2}}\hspace{15mm}\{p\}\equiv\text{frac}(p)\label{42}
\end{align}
facilitated by the property that $H^k_{a}$ is independent of $\bar{z}$ (and $t$) in the OPE limit. This mode-expansion is inverted to recover the modes as following \cite{Saha:2023hsl}:
\begin{align}
H^k_{a;n}=\frac{1}{2\pi j}\oint\limits_{C_u^\prime}d\hat{z}\text{ }\hat{z}^{n+\frac{1-k}{2}}\text{ }H^k_a(t,\hat{z},\hat{\bar{z}})
\end{align}
where $C_u^\prime$ is the counterclockwise contour on a $y=ax+b$ plane in the $1+2$D Carrollian space-time that encloses the entire upper half plane $t>0$ as well as the line $t=0$.

\medskip

Finally, we shall now find the algebra of the modes $H^k_{a;n}$. This will be (part of) the symmetry algebra manifest at the level of the OPEs of a CarrCFT that contains the field $S^+_2$. For that, we first compute the following commutator:
\begin{align*}
\left[H^k_{a;n}\text{ },\text{ }H^l_{b}(\mathbf{x}_p)\right]&=[\frac{1}{2\pi j}\oint\limits_{C_u^\prime}d\hat{z}\text{ }\hat{z}^{n+\frac{1-k}{2}}\text{ }H^k_a(t,\hat{z},\hat{\bar{z}})\text{ },\text{ }H^l_b(\mathbf{x}_p)]\\
&=\frac{1}{2\pi j}\oint\limits_{C_u}d\hat{z}\text{ }\hat{z}^{n+\frac{1-k}{2}}\text{ }\hat{\mathcal{T}}\left[ H^k_a(t_p^+,\hat{z},\hat{\bar{z}})H^l_b(t_p,\vec{x}_p)-H^k_a(t_p^-,\hat{z},\hat{\bar{z}})H^l_b(t_p,\vec{x}_p)\right]\\
&=\lim\limits_{\epsilon\rightarrow0^+}-i\left[a(l+1)-b(k+1)\right]\frac{1}{2\pi j}\oint\limits_{C_u}d\hat{z}\text{ }\hat{z}^{n+\frac{1-k}{2}}\text{ }\frac{H^{k+l-1}_{a+b}(\mathbf{x}_p)}{\hat{z}-z_p-j\epsilon0^+}-0\\
&=-i\left[a(l+1)-b(k+1)\right]{z}_p^{n+\frac{1-k}{2}}H^{k+l-1}_{a+b}(\mathbf{x}_p)
\end{align*} 
where the contour $C_u$ encloses only the upper half plane $t>0$ but not the $t=0$ line; passing from the $C^\prime_u$ to the $C_u$ is valid only in the $\epsilon\rightarrow0^+$ limit. Clearly, the $C_u$ contour does not enclose the singularity at $\hat{z}=z_p+j\epsilon0^-$ coming from the $H^k_a(t_p^-,\hat{z})H^l_b(t_p,z_p)$ term; so it has no contribution to the above commuator.

\medskip

Next, using the mode-expansions \eqref{42} for $H^{l}_{b}(z_p)$ and $H^{k+l-1}_{a+b}(z_p)$ and comparing the powers of $z_p$ on both h.s. of the above commutator, we get the following mode-algebra:
\begin{align}
i\left[H^k_{a;n}\text{ },\text{ }H^l_{b;m}\right]=\left[a(l+1)-b(k+1)\right]H^{k+l-1}_{a+b;n+m}\label{43}
\end{align}
or, in terms of the relabeled modes from \eqref{41}: $w^p_{a;n}\equiv \frac{i}{2} H^{2p-3}_{a;n}$ with $p+n\in\mathbb{Z}$ as:
\begin{align}
\left[w^p_{a;n}\text{ },\text{ }w^q_{b;m}\right]=\left[a(q-1)-b(p-1)\right]w^{p+q-2}_{a+b;n+m}
\end{align}
confirming that the (wedge sub-algebra of) $w_{1+\infty}$ Kac-Moody algebra indeed arises as the algebra of the modes from the OPEs of a CarrCFT containing the field $S^+_2$. 

\medskip

Let us have a closer look into the algebra \eqref{43} or equivalently the `OPE' \eqref{40}. As discussed earlier, the existence of the fields $S^+_1$ and $S^+_0$ is a universal feature of any $1+2$D CarrCFT since they are constructed purely in terms of the Carrollian EM tensor. The `modes' $H^1_a$ (or $w^2_a$) and $H^0_b$ (or $w^\frac{3}{2}_b$) are their respective unique signatures. In \cite{Saha:2023hsl}, from the mutual `OPE's (that remains unchanged even under the modified first assumption) of these five `modes', the symmetry algebra was derived to be the Kac-Moody extension of the $\overline{\text{sl}(2,\mathbb{R})}$ algebra with an abelian super-translation ideal. This is thus the `universal' sub-algebra of the symmetry algebra \eqref{43}. The Carrollian conformal modes $H^1_{a;n}$ generate the $\hat{\overline{\text{sl}(2,\mathbb{R})}}$ sub-algebra and the modes $H^0_{\pm\frac{1}{2};m}$ , the ideal. On the space of the Carrollian quantum fields, the three zero-modes $H^1_{a;0}$ generate the three $\overline{\text{sl}(2,\mathbb{R})}$ Lorentz transformations while the four modes $H^0_{\pm\frac{1}{2};\pm\frac{1}{2}}$ inflict the ${\text{isl}(2,\mathbb{C})}$ Poincar\'e translations.

\medskip

The special `OPE's involving the `universal modes' $H^1_a$ contain some information on representation theoretic properties of the tower of fields $S^+_k$ or, rather, their signature `modes' $H^k_a$. From the general `OPE' \eqref{40}, we readily find that:
\begin{align}
iH^1_{\pm1}H^k_{b}\sim\lim\limits_{\epsilon\rightarrow0^+}[\pm(k+1)-2b]\text{ }\frac{H^{k}_{b\pm1}}{z-z_p-j\epsilon0^+}\hspace{2.5mm};\hspace{2.5mm}iH^1_{0}H^k_{b}\sim\lim\limits_{\epsilon\rightarrow0^+}-2b\text{ }\frac{H^{k}_{b}}{z-z_p-j\epsilon0^+}
\end{align}
implying that the $(k+2)$ `modes' $H^k_{b}$ transform under the $(k+2)$-dimensional representation of the group $\overline{\text{SL}(2,\mathbb{R})}$. Consistently, the $\overline{\text{SL}(2,\mathbb{R})}$ generator `modes' $H^1_a$ transform under the 3-dimensional adjoint-representation of the said group.

\medskip

We recall that there is another `universal' Carrollian conformal field $T$, built out of the Carrollian EM tensor, that can be treated as a local field simultaneously with $S^+_0$ and $S^+_1$ \cite{Saha:2023hsl}. Since the $T\Phi$ OPEs \eqref{8} have only meromorphic pole-singularities but are anti-analytic, these have the same singularity structures as that of all the $S^+_k\Phi$ OPEs. So, the field $T$ and the tower of fields $S^+_k$ are mutually local.

\medskip

Even with the modified first assumption, the OPEs $TS^+_0$ and $TS^+_1$ remain unchanged from the following, as derived in \cite{Saha:2023hsl}:
\begin{align}
&iT(t,z,\bar{z})S^+_0(t_p,z_p,\bar{z}_p)\sim\lim\limits_{\epsilon\rightarrow0^+}\left[\frac{\frac{3}{2}S^+_0}{(\Delta\tilde{z}_p)^2}+\frac{\partial_{p}S^+_0}{\Delta\tilde{z}_p}\right](t_p,z_p,\bar{z_p})\\
&iT(t,z,\bar{z})S^+_1(t_p,z_p,\bar{z}_p)\sim\lim\limits_{\epsilon\rightarrow0^+}\left[\frac{S^+_1}{(\Delta\tilde{z}_p)^2}+\frac{\partial_{p}S^+_1}{\Delta\tilde{z}_p}-\frac{t-t_p}{2}\text{ }\frac{S^+_0}{(\Delta\tilde{z}_p)^2}\right](t_p,z_p,\bar{z_p})\nonumber
\end{align}
while the $S^+_0T$ and the $S^+_1T$ OPEs were readily obtained from the above by using the OPE commutativity property and remembering the conditions \eqref{6}, \eqref{7}, \eqref{15}, \eqref{16} and \eqref{44} to be:
\begin{align}
&iS^+_0(\mathbf{x})T(\mathbf{x}_p)\sim\lim\limits_{\epsilon\rightarrow0^+}\left[(\bar{z}-\bar{z}_p)\left(\frac{\frac{1}{2}\partial_{p}\bar{\partial}_pS^+_0}{(\Delta\tilde{z}_p)}+\frac{\frac{3}{2}\bar{\partial}_{p}S^+_0}{{(\Delta\tilde{z}_p)}^2}\right)+\frac{\frac{3}{2}S^+_0}{{(\Delta\tilde{z}_p)}^2}+\frac{\frac{1}{2}\partial_{p}S^+_0}{(\Delta\tilde{z}_p)}\right](\mathbf{x}_p)\\
&i\left\{S_1^+-(t-t_p)S^+_0\right\}(\mathbf{x})T(\mathbf{x}_p)\sim\lim\limits_{\epsilon\rightarrow0^+}\left[(\bar{z}-\bar{z}_p)^2\text{ }\frac{\frac{1}{2}\bar{\partial}^2_pS^+_1}{{(\Delta\tilde{z}_p)}^{2}}+(\bar{z}-\bar{z}_p)\text{ }\frac{\bar{\partial}_pS^+_1}{{(\Delta\tilde{z}_p)}^{2}}+\frac{S^+_1}{{(\Delta\tilde{z}_p)}^{2}}\right](\mathbf{x}_p)\nonumber
\end{align}
On the other hand, the $TT$ OPE was derived to be \cite{Saha:2023hsl}:
\begin{align}
&iT(\mathbf{x})T(\mathbf{x}_p)\sim\lim\limits_{\epsilon\rightarrow0^+}\left[\frac{-i\frac{c}{2}}{{(\Delta\tilde{z}_p)}^{4}}+\frac{2T}{(\Delta\tilde{z}_p)^2}+\frac{\partial_{z_p}T}{\Delta\tilde{z}_p}-\frac{t-t_p}{2}\left\{\frac{\frac{1}{2}\partial_{p}\bar{\partial}_pS^+_0}{(\Delta\tilde{z}_p)^2}+\frac{3\bar{\partial}_{p}S^+_0}{{(\Delta\tilde{z}_p)}^3}\right\}\right](\mathbf{x}_p)
\end{align}  
where $c$ is a constant not fixed by symmetry arguments alone.

\medskip

The (singular parts of the) general $TS^+_k$ OPEs can be completely determined by remembering the general form \eqref{8} of the $T\Phi$ OPEs and the modified first assumption to be:
\begin{align}
iT(t,z,\bar{z})S^+_k(t_p,z_p,\bar{z}_p)\sim\lim\limits_{\epsilon\rightarrow0^+}\left[\frac{\frac{3-k}{2}S^+_k}{(\Delta\tilde{z}_p)^2}+\frac{\partial_{p}S^+_k}{\Delta\tilde{z}_p}-\frac{t-t_p}{2}\text{ }\frac{S^+_{k-1}}{(\Delta\tilde{z}_p)^2}\right](t_p,z_p,\bar{z_p})\label{45}
\end{align}
that is easily checked to be consistent with the relation \eqref{28} and the restriction \eqref{29}. The $S^+_kT$ OPEs can be immediately derived from this, using the OPE commutativity property and applying these two conditions.

\medskip

Thus, together with the $S^+_0S^+_k$ and the $S^+_1S^+_k$ OPEs \eqref{30}, \eqref{31}, the $TS^+_k$ OPE \eqref{45} implies that:
\begin{align*}
\textit{each $S^+_k$ is a Carrollian conformal primary field with $\left(\bar{{\bm{\xi}}}\cdot S^+_k\right)=0=\left(\bm{\xi}\cdot S^+_k\right)$.}
\end{align*}
But, this is not the case with the field $T$. From the $S^+_0T$, $S^+_1T$ and the $TT$ OPEs above, we can immediately conclude that $T$ is not even an $\text{ISL}(2,\mathbb{C})$ quasi-primary field; it is only an $\text{SL}(2,\mathbb{C})$ or Lorentz quasi-primary \cite{Saha:2023hsl}. Moreover, both $\left(\bm{\xi}\cdot T\right)$ and $\left(\bar{{\bm{\xi}}}\cdot T\right)$ are non-zero.

\medskip

We now proceed to find the mode-algebra from the $TS^+_k$ OPE \eqref{45}. Just as the `modes' $H^k_a$, contained in the $S^+_{k(e)}$ part of the field $S^+_k$ is its unique signature, the object $T_e$ introduced in \eqref{46}, that corresponds to the 2D Celestial CFT EM tensor \cite{Kapec:2016jld}, is the unique signature of the field $T$. So, from \eqref{45}, one can easily find the following `OPE':
\begin{align}
iT_e(t,z,\bar{z})S^+_{k(e)}(t_p,z_p,\bar{z}_p)\sim\lim\limits_{\epsilon\rightarrow0^+}\left[\frac{\frac{3-k}{2} S^+_{k(e)}}{(\Delta\tilde{z}_p)^2}+\frac{\partial_{p}S^+_{k(e)}}{\Delta\tilde{z}_p}\right](t_p,z_p,\bar{z}_p)\label{47}
\end{align}
In the limit $t\rightarrow t_p^+$ and $t_p\rightarrow\infty$, this is actually a 2D Celestial CFT OPE saying that the Celestial conformally soft graviton field $S^+_{k(e)}(\infty,z,\bar{z})$ is a Celestial conformal primary \cite{Guevara:2021abz}. 

\medskip

We then note the following holomorphic mode-expansion for $T_e$ in a CarrCFT OPE \cite{Saha:2023hsl}:
\begin{align}
iT_e(z)=\sum_{n\in\mathbb{Z}}L_{n} z^{-n-2}\hspace{5mm}\Longrightarrow\hspace{5mm}L_n=\frac{1}{2\pi j}\oint\limits_{C_u^\prime}d\hat{z}\text{ }\hat{z}^{n+1}\text{ }iT_e(t,\hat{z},\hat{\bar{z}})
\end{align}
Using this and then, first the anti-holomorphic decomposition \eqref{39} for $S^+_{k(e)}$ and next the holomorphic mode-expansion \eqref{42}, we find the following commutator from the `OPE' \eqref{47} in a manner similar to the derivation of the algebra \eqref{43}:
\begin{align}
\left[L_n\text{ },\text{ }H^k_{a;m}\right]=\left(n\frac{1-k}{2}-m\right)H^{k}_{a;n+m}\hspace{2.5mm}\equiv\hspace{2.5mm}\left[L_n\text{ },\text{ }w^p_{a;m}\right]=\left[n(2-p)-m\right]w^p_{a;n+m}\label{48}
\end{align}
We also note the $[L_n,L_m]$ commutator derived in \cite{Saha:2023hsl} from the $T_eT_e$ `OPE':
\begin{align*}
\left[L_n\text{ },\text{ }L_{m}\right]=\left(n-m\right)L_{m+n}+\frac{c}{12}\left(n^3-n\right)\delta_{n+m,0}
\end{align*}
which is the holomorphic Virasoro algebra $\text{Vir}$.

\medskip

Thus, the complete symmetry algebra manifest in the OPEs of a $1+2$D CarrCFT that contains a local field $S^+_2$ obeying the relation \eqref{9} is the semi-direct product of $\text{Vir}$ and the wedge sub-algebra \cite{Pope:1991ig} of $\hat{w}_{1+\infty}$ with the semi-direct product structure given by \eqref{48}. The `universal' sub-algebra of this algebra, i.e. $\text{Vir}\ltimes\hat{\overline{\text{sl}(2,\mathbb{R})}}$ with an abelian super-translation ideal, is the OPE-level symmetry (in the holomorphic sector) of any $1+2$D CarrCFT \cite{Saha:2023hsl}.

\medskip

Unlike the Celestial CFT literature \cite{Banerjee:2020zlg,Guevara:2021abz,Strominger:2021lvk,Banerjee:2021dlm,Himwich:2021dau}, we reached the above conclusion solely from the Carrollian conformal symmetry arguments and the general properties of OPEs, under the two assumptions stated in section \ref{s3.1}, without requiring any hint from the explicit (quantum) theory of gravitation in the $1+3$D bulk AFS. Thus, our analysis is purely holographic in nature.    

\medskip

\section{An Infinity of Soft Theorems}\label{s6}
We shall now uncover, in the current framework of the $1+2$D CarrCFT, the direct connection between the existence of the infinite tower of conformally soft graviton fields, as described in \cite{Guevara:2021abz} in the context of the 2D Celestial CFT and in section \ref{s4} of this work and an infinite number of soft graviton theorems manifest as the Ward identities of large diffeomorphisms, as presented in \cite{Hamada:2018vrw,Li:2018gnc}.

\medskip

As suggested in section \ref{s3}, for a Carrollian conformal field $S^+_2$ to encode the subsubleading energetically soft graviton theorem \cite{Cachazo:2014fwa} in its Ward identity, it should obey a relation like \eqref{9}. As we have seen in sections \ref{s4} and \ref{s5}, recursive iteration of this suggestion reproduces the correct Carrollian conformal OPEs containing the conformally soft graviton OPEs \cite{Guevara:2021abz} of 2D Celestial CFT. Providing a direct justification to this suggestion is the first step towards the goal of this section. We shall closely follow the argument presented in \cite{Saha:2023hsl} that showed the relation between the source-less $1+2$D CarrCFT EM tensor Ward identities and the leading \cite{Weinberg:1965nx} and the subleading \cite{Cachazo:2014fwa} soft graviton theorems.

\medskip

We begin by noting down the $S^+_2S^+_k$ OPE \eqref{32} below: 
\begin{align*}
iS_2^+(\mathbf{x})S^+_k(\mathbf{x}_p)&\sim\lim\limits_{\epsilon\rightarrow0^+}\left[\frac{(\bar{z}-\bar{z}_p)^3}{{(\Delta\tilde{z}_p)}}\text{ }\frac{1}{2}\bar{\partial}_p^2S^+_{k+1}+\frac{(\bar{z}-\bar{z}_p)^2}{{(\Delta\tilde{z}_p)}}\text{ }(k+1)\bar{\partial}_pS^+_{k+1}+\frac{\bar{z}-\bar{z}_p}{{(\Delta\tilde{z}_p)}}\text{ }\frac{1}{2}(k+1)(k+2)S^+_{k+1}\right.\nonumber\\
&\left.+(t-t_p)\left\{\frac{(\bar{z}-\bar{z}_p)^2}{{(\Delta\tilde{z}_p)}}\text{ }\bar{\partial}_{p} S^+_k+\frac{\bar{z}-\bar{z}_p}{{(\Delta\tilde{z}_p)}}\text{ }(k+1)S^+_k\right\}+\frac{1}{2}(t-t_p)^2\text{ }\frac{\bar{z}-\bar{z}_p}{{(\Delta\tilde{z}_p)}}\text{ }S^+_{k-1}\right](\mathbf{x}_p)
\end{align*}
and recall that all the fields $S^+_k$ are Carrollian conformal primaries with $\left(\bar{{\bm{\xi}}}\cdot S^+_k\right)=0=\left(\bm{\xi}\cdot S^+_k\right)$ with dimensions $(h,\bar{h})=\left(\frac{3-k}{2},-\frac{k+1}{2}\right)$. Since valid for an infinite number of fields $S^+_k$, we postulate that the field $S^+_2$ has the following OPE with a special Carrollian conformal primary\footnote{We expect the Carrollian conformal weight $\Delta$ to be discrete unlike its Celestial counterpart $\Delta_c$. This expectation stems from the fact that a $1+2$D Carrollian conformal primary field that corresponds to a 4D bulk field describing a mass-less (hard) scattering particle must possess $\Delta=1$. Clearly, all the descendants of such a field have integer Carrollian weights.} $\Phi$ with dimensions $(h,\bar{h})$ and $\left(\bar{{\bm{\xi}}}\cdot \Phi\right)=0=\left(\bm{\xi}\cdot \Phi\right)$:
\begin{align}
iS_2^+(\mathbf{x})\Phi(\mathbf{x}_p)&\sim\lim\limits_{\epsilon\rightarrow0^+}\left[\frac{(\bar{z}-\bar{z}_p)^3}{{(\Delta\tilde{z}_p)}}\text{ }\frac{1}{2}\bar{\partial}_p^2\Phi_1-\frac{(\bar{z}-\bar{z}_p)^2}{{(\Delta\tilde{z}_p)}}\text{ }2\bar{h}\bar{\partial}_p\Phi_1+\frac{\bar{z}-\bar{z}_p}{{(\Delta\tilde{z}_p)}}\text{ }\frac{1}{2}(2\bar{h})(2\bar{h}-1)\Phi_1\right.\nonumber\\
&\left.+(t-t_p)\left\{\frac{(\bar{z}-\bar{z}_p)^2}{{(\Delta\tilde{z}_p)}}\text{ }\bar{\partial}_{p} \Phi-\frac{\bar{z}-\bar{z}_p}{{(\Delta\tilde{z}_p)}}\text{ }2\bar{h}\Phi\right\}+\frac{1}{2}(t-t_p)^2\text{ }\frac{\bar{z}-\bar{z}_p}{{(\Delta\tilde{z}_p)}}\text{ }\dot{\Phi}\right](\mathbf{x}_p)\label{49}
\end{align} 
with the unique local Carrollian field $\Phi_1$ satisfying $\dot{\Phi}_1\sim\Phi$. It is important to note that the OPE of the local field $\Phi_1$ must not be completely determinable in terms of that of the field $\Phi$; more precisely, it should hold that:
\begin{align*}
\Phi_1(t,z,\bar{z})\sim\int\limits_{-\infty}^t dt^\prime\text{ }\Phi(t^\prime,z,\bar{z})+\text{(terms with temporal step-function factor)}
\end{align*}
as happens for the fields $S^+_k$. 

\medskip

Remembering the decomposition \eqref{33}, we first collect the $S^+_{2(e)}\Phi$ `OPE' as:
\begin{align*}
iS_{2(e)}^+(t,z,\bar{z})\Phi(\mathbf{x}_p)\sim\lim\limits_{\epsilon\rightarrow0^+}&\left[\frac{(\bar{z}-\bar{z}_p)^3}{{(\Delta\tilde{z}_p)}}\text{ }\frac{1}{2}\bar{\partial}_p^2\Phi_1-\frac{(\bar{z}-\bar{z}_p)^2}{{(\Delta\tilde{z}_p)}}\text{ }2\bar{h}\bar{\partial}_p\Phi_1+\frac{\bar{z}-\bar{z}_p}{{(\Delta\tilde{z}_p)}}\text{ }\frac{1}{2}(2\bar{h})(2\bar{h}-1)\Phi_1\right.\nonumber\\
&\left.-t_p\left\{\frac{(\bar{z}-\bar{z}_p)^2}{{(\Delta\tilde{z}_p)}}\text{ }\bar{\partial}_{p} \Phi-\frac{\bar{z}-\bar{z}_p}{{(\Delta\tilde{z}_p)}}\text{ }2\bar{h}\Phi\right\}+\frac{t_p^2}{2}\text{ }\frac{\bar{z}-\bar{z}_p}{{(\Delta\tilde{z}_p)}}\text{ }\dot{\Phi}\right](\mathbf{x}_p)
\end{align*}
that is the Carrollian conformal counterpart of the subsubleading conformally soft graviton OPE with a 2D Celestial conformal primary as in \cite{Banerjee:2021cly}, since clearly, $\Delta_{\Phi_1}=\Delta_{\Phi}-1$ and $m_{\Phi_1}=m_\Phi$. This observation is consistent with the earlier interpretation in section \ref{s5} of the object $S^+_{2(e)}(\infty,z,\bar{z})$ as the Celestial conformally soft graviton field $H^{-1}(z,\bar{z})$ in \cite{Guevara:2021abz}. This `OPE' is further decomposed according to the anti-holomorphic `mode-expansion' \eqref{39} as:
\begin{align}
&iH^2_{-\frac{3}{2}}(z)\Phi(\mathbf{x}_p)\sim\lim\limits_{\epsilon\rightarrow0^+}\frac{3\bar{\partial}_p^2\Phi_1}{(\Delta\tilde{z}_p)}\hspace{3.5mm};\hspace{3.5mm}iH^2_{-\frac{1}{2}}(z)\Phi(\mathbf{x}_p)\sim\lim\limits_{\epsilon\rightarrow0^+}\frac{3\bar{z}_p\bar{\partial}_p^2\Phi_1+4\bar{h}\bar{\partial}_p\Phi_1+2t_p\bar{\partial}_p\Phi}{(\Delta\tilde{z}_p)}\nonumber\\
&iH^2_{\frac{1}{2}}(z)\Phi(\mathbf{x}_p)\sim\lim\limits_{\epsilon\rightarrow0^+}\frac{3\bar{z}^2_p\bar{\partial}_p^2\Phi_1+8\bar{h}\bar{z}_p\bar{\partial}_p\Phi_1+2\bar{h}(2\bar{h}-1)\Phi_1+4\bar{z}_p t_p\bar{\partial}_p\Phi+4\bar{h}t_p\Phi+t_p^2\dot{\Phi}}{(\Delta\tilde{z}_p)}\label{51}\\
&iH^2_{\frac{3}{2}}(z)\Phi(\mathbf{x}_p)\sim\lim\limits_{\epsilon\rightarrow0^+}3\frac{\bar{z}^3_p\bar{\partial}_p^2\Phi_1+4\bar{h}\bar{z}^2_p\bar{\partial}_p\Phi_1+2\bar{h}(2\bar{h}-1)\bar{z}_p\Phi_1+2\bar{z}^2_p t_p\bar{\partial}_p\Phi+4\bar{h}t_p\bar{z}_p\Phi+t_p^2\bar{z}_p\dot{\Phi}}{(\Delta\tilde{z}_p)}\nonumber
\end{align} 
where it is understood that the fields on the r.h.s. are at $\mathbf{x}_p$. Finally, the holomorphic mode-expansion of the `modes' $H^2_k$ are given by \eqref{42} as:
\begin{align*}
H^2_a(z)=\sum_{n\in\mathbb{Z}+\frac{1}{2}}H^2_{a;n} z^{-n-\frac{1}{2}}
\end{align*}
that implies the following transformation generated by, e.g. the modes $H^2_{-\frac{1}{2};n}$ , from \eqref{51}:
\begin{align}
i\left[H^2_{-\frac{1}{2};n}\text{ },\text{ }\Phi(\mathbf{x}_p)\right]={z}_p^{n-\frac{1}{2}}\left[3\bar{z}_p\bar{\partial}_p^2\Phi_1+4\bar{h}\bar{\partial}_p\Phi_1+2t_p\bar{\partial}_p\Phi\right](\mathbf{x}_p)\label{52}
\end{align}
derived using the CarrCFT OPE $\longleftrightarrow$ commutator prescription developed in \cite{Saha:2023hsl}. 

\medskip

Thus, the four Carrollian conformal modes $H^2_{a;\frac{1}{2}}$ generate four global transformations on the quantum field $\Phi$, that can be read off of the numerators of the `OPE's \eqref{51}. These are the four global transformations generated by the Carrollian conformal field $S^+_2$. Consequently, the existence of the field $S^+_2$ in the theory demands that the CarrCFT correlators must be invariant under these four global transformations, in addition to the ten global Poincar\'e constraints imposed by the three `universal' generators $S^+_1$, $S^+_0$ and $T$. 

\medskip

The most striking feature of the transformations generated by $S^+_2$ is their non-locality in time (but locality in space): they involve a time-integral of the original primary field. This is consistent with the conclusion of an Einstein-gravity analysis in \cite{Freidel:2021dfs} that the subsubleading soft graviton theorem arises as a consequence of conservation of a spin-2 charge generating a non-local space-time symmetry at null infinity; these symmetry transformations are also non-local only in (retarded-)time.

\medskip

That in the $\mathcal{O}\left((t-t_p)^0\right)$ terms of the $S^+_2\Phi$ OPE \eqref{49}, it is indeed $\bar{h}$ that appears instead of $h$ can be checked by considering e.g., a Jacobi identity involving the primary $\Phi$ and the two Carrollian conformal modes $H^2_{-\frac{1}{2};\frac{1}{2}}$ and $H^0_{\frac{1}{2};-\frac{1}{2}}$. There is no doubt about the terms linear and quadratic in $(t-t_p)$ that are already fixed by the restriction \eqref{9}. Also, by considering another Jacobi identity with $H^0_{\frac{1}{2};-\frac{1}{2}}$ replaced by $H^0_{\frac{1}{2};\frac{1}{2}}$, it becomes apparent that the form of the $\mathcal{O}\left((t-t_p)^0\right)$ terms in the OPE \eqref{49} must be modified when the primary $\Phi$ has non-zero $\left(\bar{{\bm{\xi}}}\cdot \Phi\right)$ and/or $\left({{\bm{\xi}}}\cdot \Phi\right)$ or $\Phi$ is a non-primary. An important example of the second case is the $S^+_2T$ OPE whose $\mathcal{O}\left((t-t_p)^0\right)$ terms are very different from those in \eqref{49}. In this work, we do not consider the case of the general primaries because only the Carrollian primaries with $\left(\bar{{\bm{\xi}}}\cdot \Phi\right)=0=\left({{\bm{\xi}}}\cdot \Phi\right)$ can describe mass-less scattering in the bulk AFS \cite{Saha:2023hsl}.

\medskip

Restoring the temporal step-function factor, the $S^+_2$ Ward identity corresponding to the OPE \eqref{49} is given by: 
\begin{align}
i\langle S_2^+(t,z,\bar{z}) X\rangle=\sum\limits_{p=1}^n\theta&(t-t_p)\left[\frac{(\bar{z}-\bar{z}_p)^3}{{z-z_p}}\text{ }\frac{1}{2}\bar{\partial}_p^2\partial_{t_p}^{-1}-\frac{(\bar{z}-\bar{z}_p)^2}{{z-z_p}}\text{ }2\bar{h}_p\bar{\partial}_p\partial_{t_p}^{-1}+\frac{\bar{z}-\bar{z}_p}{{z-z_p}}\text{ }\bar{h}_p(2\bar{h}_p-1)\partial_{t_p}^{-1}\right.\nonumber\\
+&\left.(t-t_p)\left\{\frac{(\bar{z}-\bar{z}_p)^2}{z-{z}_p}\bar{\partial}_{p}-2\bar{h}_p\frac{\bar{z}-\bar{z}_p}{z-{z}_p}\right\}+\frac{(t-t_p)^2}{2}\text{ }\frac{\bar{z}-\bar{z}_p}{z-{z}_p}\partial_{t_p}\right]\langle X\rangle\label{50}
\end{align}
with $X$ being a string of Carrollian conformal primaries, all with $\left(\bar{{\bm{\xi}}}\cdot \Phi\right)=0=\left(\bm{\xi}\cdot \Phi\right)$. The terms regular in $z-z_p$ all vanish because we demand that the $\langle S_2^+(t,z,\bar{z})X\rangle$ correlator be finite whenever $z\neq z_p$; in particular, this correlator must be finite when $z=\infty$ (remembering that the CarrCFT primary $S^+_2$ has $h>0$). The complete argument is analogous to the ones elaborated in \cite{Saha:2023hsl} for the finite-ness of the correlators $\langle S_1^+X\rangle$, $\langle S_0^+X\rangle$ and $\langle TX\rangle$. Since the integral operator $\partial_{t_p}^{-1}$ clearly changes $(\Delta,m)\rightarrow(\Delta-1,m)$, the $S^+_{2(e)}$ part of this Ward identity is the Carrollian conformal version of the positive-helicity subsubleading conformally soft graviton Ward identity in \cite{Pate:2019lpp}.

\medskip

In \cite{Saha:2023hsl}, it was shown how to recover the $1+3$D bulk AFS leading \cite{Weinberg:1965nx} and the subleading \cite{Cachazo:2014fwa} energetically soft graviton theorems by simply taking a temporal Fourier transformation \cite{Strominger:2014pwa,Pasterski:2015tva,Strominger:2017zoo} of the $S^+_1$ Ward identity \eqref{10} and then imposing an energetically soft limit only for the field $S^+_1$. To perform the temporal Fourier transformation, opposite phase conventions were required \cite{Donnay:2022wvx} for the Carrollian conformal primaries with $\Delta=1$ that were to describe the outgoing or incoming bulk AFS mass-less particles in null-momentum space \cite{Donnay:2022aba,Donnay:2022wvx,Nguyen:2023vfz}; this convention was chosen in \cite{Saha:2023hsl} as (with $\omega\geq0$):
\begin{align}
\tilde{\Phi}_{\text{out}}(\omega,z,\bar{z})=\frac{1}{2\pi}\int\limits_{-\infty}^{\infty}dt \text{ }e^{-i\omega t}{\Phi}(t,z,\bar{z})\hspace{5mm}\text{ and }\hspace{5mm}\tilde{\Phi}_{\text{in}}(\omega,z,\bar{z})=\frac{1}{2\pi}\int\limits_{-\infty}^{\infty}dt \text{ }e^{i\omega t}{\Phi}(t,z,\bar{z})\label{56}
\end{align}

\medskip

We now temporal Fourier transform the $S^+_2$ Ward identity \eqref{50} according to the above convention while choosing the outgoing convention for $S^+_2$ , complexify the energy $\omega$ of the field $S^+_2$ , set all $\Delta_p=1$ , $m_p=s_p$ (the helicity of the bulk mass-less particle) and finally, take the energetically soft $\omega\rightarrow0$ limit to obtain\footnote{modulo the zero-mode problem of $\partial_{t_p}^{-1}$ .}:
\begin{align}
\lim\limits_{\omega\rightarrow0}&i\left\langle \tilde{S}_2^+(\omega,z,\bar{z}) \tilde{X}_{\text{out}}\tilde{X}_{\text{in}}\right\rangle=\lim\limits_{\omega\rightarrow0}-\sum\limits_{p\in\text{all}}\left\{\frac{\omega_p}{\omega^3}\frac{\bar{z}-\bar{z}_p}{z-z_p}\epsilon_p+\frac{1}{\omega^2}\frac{(\bar{z}-\bar{z}_p)^2\bar{\partial}_{p}+(\bar{z}-\bar{z}_p)\left(s_p+\omega_p\partial_{\omega_p}\right)}{{z}-{z}_p}\right.\nonumber\\
&\hspace{19mm}\left.+\frac{1}{\omega}\frac{\epsilon_p}{2\omega_p}\frac{{\left[(\bar{z}-\bar{z}_p)^2\bar{\partial}_{p}+(\bar{z}-\bar{z}_p)\left(s_p+\omega_p\partial_{\omega_p}\right)\right]}^2}{(z-z_p)(\bar{z}-\bar{z}_p)}+\mathcal{O}(\omega^0)\right\}\left\langle  \tilde{X}_{\text{out}}\tilde{X}_{\text{in}}\right\rangle\label{57}
\end{align}
with $\epsilon_p=+1$ for outgoing and $\epsilon_p=-1$ for incoming fields. As expected, the r.h.s. is recognized as the soft factorization and energetically (holomorphic\footnote{Inside a Carrollian correlator, the relation $\partial^2_tS^+_2= S^+_0$ is translated into $\omega^2\tilde{S}_2^+=-\tilde{S}_0^+$ in the $\omega$-space; so, it is consistent with the fact that insertion of $\lim\limits_{\omega\rightarrow0}i\tilde{S}_0^+(\omega,z,\bar{z})$ leads to the `natural' soft-factorization \cite{Cachazo:2014fwa}.} \cite{Cachazo:2014fwa}) soft expansion of the soft factor of a $1+3$D bulk AFS mass-less scattering amplitude involving an outgoing soft external graviton up to the subsubleading order. The leading $\mathcal{O}(\frac{1}{\omega^3})$ term is the Weinberg universal soft factor \cite{Weinberg:1965nx,He:2014laa}, the subleading $\mathcal{O}(\frac{1}{\omega^2})$ term is the Cachazo-Strominger universal soft factor \cite{Cachazo:2014fwa,Kapec:2014opa} and the subsubleading $\mathcal{O}(\frac{1}{\omega})$ is a non-universal \cite{Laddha:2017ygw,Elvang:2016qvq} soft factor; its form agrees with the one in \cite{Conde:2016rom} explicitly calculated for the example of tree-level Einstein-gravity \cite{Cachazo:2014fwa}. Within the Carrollian context, this subsubleading soft factor is non-universal since it is present only in those $1+2$D CarrCFTs that contain the field $S^+_2$, unlike the two in the more leading orders \cite{Saha:2023hsl}.

\medskip

Thus, we have shown that the Ward identity of the Carrollian conformal field $S^+_2$ obeying the relation \eqref{9}, indeed encodes the bulk AFS soft graviton theorems up to the subsubleading order. In section \ref{s5}, it was also concluded that this $S^+_2$ field contains the positive-helicity subsubleading conformally soft graviton primary \cite{Pate:2019lpp,Guevara:2021abz} of the 2D Celestial CFT.   

\medskip

Inspired by the above conclusions for the field $S^+_2$, one may get tempted to draw a blind analogy for the whole infinite tower of fields $S^+_k$ containing, as shown in section \ref{s5}, the whole family of the 2D Celestial conformal primary soft gravitons of \cite{Guevara:2021abz}. We shall find out that while this infinite tower of fields $S^+_k$ indeed encodes, as Ward identities, an infinity of projected energetically soft theorems described in \cite{Hamada:2018vrw,Li:2018gnc}, the fields $S^+_{k>2}$ do not generate any additional global symmetries and hence, impose no new constraints on the CarrCFT correlators, consistent with \cite{Guevara:2021abz}.    

\medskip

We begin by rewriting the general $S^+_kS^+_l$ OPE \eqref{37} as below:
\begin{align*}
iS_k^+(\mathbf{x})S^+_{l}(\mathbf{x}_p)\sim\lim\limits_{\epsilon\rightarrow0^+}\sum_{r=0}^k
\frac{\left(t-t_p\right)^{k-r}}{(k-r)!}\sum_{m=0}^r\frac{\left(\bar{z}-\bar{z}_p\right)^{m+1}}{{(\Delta\tilde{z}_p)}}\text{ }\frac{(l+1)_{r-m}}{(r-m)!\cdot m!}\text{ }\bar{\partial}^m_p S^+_{r+l-1}(\mathbf{x}_p)
\end{align*}
and recall that dimensions of the field $S^+_l$ are $(h,\bar{h})=\left(\frac{3-l}{2},-\frac{1+l}{2}\right)$. In exact similarity with the case of $S^+_2$, we postulate the following $S^+_k\Phi$ OPE:
\begin{align}
iS_k^+(\mathbf{x})\Phi(\mathbf{x}_p)\sim\lim\limits_{\epsilon\rightarrow0^+}\sum_{r=0}^k
\frac{\left(t-t_p\right)^{k-r}}{(k-r)!}\sum_{m=0}^r\frac{\left(\bar{z}-\bar{z}_p\right)^{m+1}}{{(\Delta\tilde{z}_p)}}\text{ }\frac{(-2\bar{h})_{r-m}}{(r-m)!\cdot m!}\text{ }\bar{\partial}^m_p \Phi_{r-1}(\mathbf{x}_p)\label{53}
\end{align}
for a special Carrollian conformal primary $\Phi$ with dimensions $(h,\bar{h})$ and with $\left(\bar{{\bm{\xi}}}\cdot \Phi\right)=0=\left({{\bm{\xi}}}\cdot \Phi\right)$. The unique local fields $\left\{\Phi_r\right\}$ satisfy:
\begin{align*}
\partial_t^r\Phi_r\sim\Phi \text{ ($r\geq1$)}\hspace{5mm},\hspace{5mm}\Phi_0=\Phi\hspace{5mm}\text{and}\hspace{5mm}\Phi_{-1}=\dot{\Phi}
\end{align*}

\medskip

Following the discussion on $S^+_2$, we can extract the `OPE's $H^k_a(z)\Phi(\mathbf{x}_p)$ analogous to \eqref{51}, by appealing to the decomposition \eqref{33} and \eqref{39} and from there, derive the space-time transformations inflicted by the Carrollian conformal modes $H^k_{a;n}$ on the quantum fields, just like \eqref{52}. It is clear that all of these transformations are non-local in time but spatially local as they involve time-integrals of various order of the original primary field. For a discussion on the Einstein-gravity dual of these transformations, see \cite{Freidel:2021ytz}.

\medskip

We are now in a position to conclude that the infinite number of $S^+_k\Phi$ OPEs will thus generate an infinite number of global space-time symmetry transformations under all of which the CarrCFT correlators must be invariant, just as we did for the $S^+_2$ case. But that is not the case! 

\medskip

To show this, let us start with the case of $S^+_3$. The transformations that its modes generate is derived as $\left[H^3_{a;n}\text{ },\text{ }\Phi(\mathbf{x}_p)\right]$ via the OPE $\longleftrightarrow$ commutator prescription starting from the $S^+_3\Phi$ OPE. But, from the symmetry algebra \eqref{43}, we see that:
\begin{align*}
i\left[H^2_{a;n}\text{ },\text{ }H^2_{b;m}\right]=3\left(a-b\right)H^3_{a+b;n+m}
\end{align*}
Thus, the transformation $\left[H^3_{a;n}\text{ },\text{ }\Phi(\mathbf{x}_p)\right]$ can be directly found out using a Jacobi identity involving the field $\Phi$ and two appropriate modes $H^2_{r;l}$ and $H^2_{s;m}$ and the knowledge of the transformations $\left[H^2_{b;k}\text{ },\text{ }\Phi(\mathbf{x}_p)\right]$, without any need to learn the $S^+_3\Phi$ OPE at all. We can iterate this process to extract the transformations generated by the modes of the field $S^+_{k>2}$ from the knowledge of the transformations inflicted by the modes of $S^+_2$ and $S^{+}_{k-1}$ without ever appealing to the $S^+_k\Phi$ OPE. Hence by induction, all we require to find the transformations generated by the tower of fields $S^+_{k>2}$ is the knowledge of only the transformations that $S^+_2$ generate. Thus, the seemingly infinite number of global symmetries are not independent at all from the four generated by $S^+_2$. Thus, the correlators in such a CarrCFT are subject to a merely four additional global symmetry constraints in addition to the ten `universal' Poincar\'e constraints. This conclusion resonates with the observation in section \ref{s4} that in a CarrCFT containing a field $S^+_2$ obeying \eqref{9}, an infinite tower of fields $S^+_{k\geq3}$ are required to automatically exist to render the $S^+_2S^+_{k-1}$ OPEs consistent.

\medskip

Since the temporal Fourier transformation of a $1+2$D position-space CarrCFT correlator of primaries all with $\left(\bar{{\bm{\xi}}}\cdot \Phi\right)=0=\left({{\bm{\xi}}}\cdot \Phi\right)$ \cite{Saha:2023hsl} and $\Delta=1$ \cite{Donnay:2022aba,Donnay:2022wvx,Nguyen:2023vfz} gives the $1+3$D bulk AFS null-momentum space $S$-matrix \cite{Donnay:2022wvx}, the above discussion implies that no new global symmetry constraint besides the ten `universal' Poincar\'e plus the four generated by the field $S^+_2$ is imposed on this $S$-matrix by the tower of fields $S^+_{k\geq3}$. The 2D Celestial CFT counterpart of this statement is proved in \cite{Guevara:2021abz}. 

\medskip

We shall now finally establish that the infinite tower of Carrollian conformal primaries $S^+_k$ does indeed imply the existence of an infinity of projected (energetically) soft graviton theorems \cite{Hamada:2018vrw,Li:2018gnc} as their Ward identities. 

\medskip

Restoring the temporal step-function, the Ward identity corresponding to the CarrCFT OPE \eqref{53} is given as the following:
\begin{align}
i\langle S_k^+(t,z,\bar{z}) X\rangle=\sum_{p=1}^n\theta(t-t_p)\sum_{r=0}^k
\frac{\left(t-t_p\right)^{k-r}}{(k-r)!}\sum_{m=0}^r\frac{\left(\bar{z}-\bar{z}_p\right)^{m+1}}{{z-z_p}}\text{ }\frac{(-2\bar{h}_p)_{r-m}}{(r-m)!\cdot m!}\text{ }\bar{\partial}^m_p \partial_{t_p}^{1-r}\langle X\rangle\label{55}
\end{align}
up to a $(k-3)$-th degree polynomial in $z$ that can not be fixed by symmetry considerations alone. Obviously, it will now receive the same treatment as the $S^+_2$ Ward identity \eqref{50}. For that, we first note the following identity:
\begin{align}
\int\limits_{-\infty}^{\infty}dt \text{ }e^{-i\omega t}\theta(t-t_p)\text{ }\frac{(t-t_p)^s}{s!}=\lim\limits_{a\rightarrow0^+}\lim\limits_{b\rightarrow0^+}\frac{e^{-i\omega t_p}}{\left[i\omega+(b-ia)\right]^{s+1}}\label{54}
\end{align}
where $\omega$ is a complex quantity. 

\medskip

Following the convention \eqref{56}, we now temporal Fourier transform the $S^+_k$ Ward identity \eqref{55} choosing the outgoing convention for $S^+_k$ , set all $\Delta_p=1$, $m_p=s_p$, use the identity \eqref{54} (and explicitly put the limits for $a,b$) and finally impose the energetically soft $\omega\rightarrow0$ limit to obtain the following schematic Laurent series around $\omega=0$:
\begin{align*}
\lim\limits_{\omega\rightarrow0}&i^{k+1}\left\langle \tilde{S}_k^+(\omega,z,\bar{z}) \tilde{X}_{\text{out}}\tilde{X}_{\text{in}}\right\rangle=\lim\limits_{\omega\rightarrow0}\left[\frac{F^{(0)}}{\omega^{k+1}}+\frac{F^{(1)}}{\omega^{k}}+\frac{F^{(2)}}{\omega^{k-1}}+\ldots+\frac{F^{(k)}}{\omega}+\mathcal{O}(\omega^0)\right] \left\langle\tilde{X}_{\text{out}}\tilde{X}_{\text{in}}\right\rangle
\end{align*}
where $F^{(0)}$ is the Weinberg leading soft factor \cite{Weinberg:1965nx,He:2014laa}, $F^{(1)}$ is the Cachazo-Strominger subleading soft factor \cite{Cachazo:2014fwa,Kapec:2014opa} and $F^{(2)}$ is the subsubleading soft factor \cite{Cachazo:2014fwa,Conde:2016rom} that appear in \eqref{57}; $F^{(k\geq3)}$ are the next order soft factors. A soft factor $F^{(k)}$ has $(z-z_p)$ in the denominator and a $(k+1)$-th degree polynomial in $(\bar{z}-\bar{z}_p)$ in the numerator. It is possible (but tedious) to derive its explicit form following our derivation of $F^{(2)}$. From the explicit examples of the $S^+_0,S^+_1,S^+_2$ cases, it is clear that the soft factor $F^{(k)}$ first appears from the Ward identity of the field $S^+_k$ at the order $\mathcal{O}(\frac{1}{\omega})$. In this manner, an infinite number of projected energetically soft graviton theorems \cite{Hamada:2018vrw,Li:2018gnc} arises from the Ward identities of the infinite number of Carrollian conformal primaries $S^+_k$ that contain the 2D Celestial conformal primary soft gravitons $H^{1-k}$ of \cite{Guevara:2021abz}. Furthermore, the undetermined terms polynomial in $z$ in the $\langle S^+_kX\rangle$ Ward identity correspond to the homogeneous part of the graviton amplitude that are projected out to obtain the infinite-order soft factorization \cite{Hamada:2018vrw,Li:2018gnc}.

\medskip

\section{Discussion}\label{s7}
Building on the direct relation between the EM tensor Ward identities of a $1+2$D source-less CarrCFT on a flat Carrollian background (with $\mathbb{R}\times S^2$ topology) and the universal leading \cite{Weinberg:1965nx} and the subleading \cite{Cachazo:2014fwa} soft graviton theorems that was uncovered in \cite{Saha:2023hsl}, in this work we investigated how the non-universal \cite{Laddha:2017ygw,Elvang:2016qvq} subsubleading soft graviton theorem \cite{Cachazo:2014fwa} can be holographically encoded into a CarrCFT if at all.

\medskip

We found out that in addition to the following three universal local generator fields in any generic $1+2$D CarrCFT: $S^+_0$ (and its non-local shadow $S^-_0$) whose Ward identity contains the leading soft graviton theorems and $S^+_1$ and $T$ encoding the subleading ones (of both helicities) \cite{Saha:2023hsl}, a local Carrollian conformal field $S^+_2$ must be postulated to exist in the theory such that it obeys the relation \eqref{9}, for capturing the positive-helicity subsubleading soft graviton theorem. To avoid the ambiguity associated with double soft limits of opposite helicities \cite{Klose:2015xoa}, here we have refrained from looking into the case of the negative-helicity non-universal subsubleading soft theorem.

\medskip

After introducing the local field $S^+_2$ in the theory, we attempted to construct the mutual OPEs of the three fields $S^+_0,S^+_1,S^+_2$ using only the general forms of the CarrCFT OPEs \eqref{5} and \eqref{14}, derived completely from Carrollian symmetry arguments in \cite{Saha:2023hsl}, and the OPE commutativity property. For this method to work, we needed to assume that there is no time-independent local field with negative scaling dimension $\Delta$ in the theory with the Identity being the unique time-independent field with dimensions $(\Delta,m)=(0,0)$.

\medskip

In \cite{Zamolodchikov:1989mz} and in \cite{Zamolodchikov:1985wn}, it was assumed that no local field with negative holomorphic weight $h$ exists in the 2D Euclidean (chiral) CFT, with the unique field having $h=0$ being the Identity, to completely determine the singular parts of respectively the $TT$ OPE and the $JJ$ OPEs using only general conformal symmetry principles and the OPE commutativity property. $T$ is the holomorphic EM tensor and $J$ represents an additional symmetry generating primary with $h\in\{\frac{1}{2},1,\frac{3}{2},2,\frac{5}{2},3\}$. The above mentioned CarrCFT assumption is similar in spirit to this one. Since, $h>0$ is the unitary bound in a 2D chiral CFT, we wonder if the non-existence of time-independent local fields with $\Delta<0$ is analogously a unitarity condition for a $1+2$D CarrCFT. It will be very interesting to establish or dismiss this speculation. 

\medskip

Proceeding to construct the OPEs using only the general symmetry arguments under this assumption, we discovered that for the $S^+_2S^+_2$ OPE to be consistent, there must automatically exist another local primary $S^+_3$ in the theory, obeying the relation \eqref{24}. Similarly, it was found that another local field $S^+_4$, satisfying \eqref{58} must automatically exist to render the $S^+_2S^+_3$ OPE consistent. Iterating the algorithm, it could be seen that an OPE $S^+_2S^+_k$ is consistent if a local field $S^+_{k+1}$ is automatically present in the theory, that obeys the condition \eqref{28}. Thus, the local primary $S^+_2$, if present in the CarrCFT, generates an infinite tower of Carrollian conformal primaries $S^+_{k\geq3}$. Each of these primaries satisfies a null-state condition \eqref{29}.

\medskip

The general Carrollian conformal OPE $S^+_k(\mathbf{x})S^+_l(\mathbf{x}_p)$ was obtained as \eqref{37}. In this derivation, the Carrollian time coordinate played the central role via the condition \eqref{28}. Comparing the $\mathcal{O}(t^0t_p^0)$ terms on both sides of this OPE and then imposing the $t\rightarrow\infty$ limit, we recovered the following OPE of two conformally soft graviton primaries: $H^{1-k}(z,\bar{z})H^{1-l}(z_p,\bar{z}_p)$ of the 2D Celestial CFT \cite{Guevara:2021abz}. This Celestial OPE was obtained in \cite{Guevara:2021abz} by taking the conformally soft limit of the OPE of two conformal primary gravitons with arbitrary (Celestial) weights that was itself derived in \cite{Pate:2019lpp} for the specific case of the bulk tree-level (linearized) Einstein gravity. On the contrary, the Carrollian derivation presented here needed no input from the explicit theory of quantum gravity in the bulk AFS and hence, is purely holographic. 

\medskip

Performing the Carrollian conformal mode-expansion according to the decomposition \eqref{33} and \eqref{39} of each primary $S^+_k$ with holomorphic weight $h=\frac{3-k}{2}$, we found, using the OPE $\longleftrightarrow$ commutator map developed in \cite{Saha:2023hsl}, that the symmetry algebra manifest in the CarrCFT OPEs \eqref{37} is the (wedge sub-algebra \cite{Pope:1991ig} of the) $w_{1+\infty}$ Kac-Moody algebra, in perfect agreement with the conclusion reached in \cite{Strominger:2021lvk} in the context of 2D Celestial CFT. Thus, the complete symmetry algebra at the level of the OPEs of a CarrCFT containing the field $S^+_2$ is the semi-direct product of the (chiral) Virasoro algebra and the wedge sub-algebra of $\hat{w}_{1+\infty}$ ; the semi-direct product structure is given by \eqref{48} arising from the $TS^+_k$ OPEs \eqref{45}.

\medskip

It is important to recall that we did not rely on a (Carrollian) perturbative analysis to obtain the OPEs (under the assumptions mentioned in section \ref{s3.1} though) from which the above symmetry algebra was derived. The starting point of the construction of these OPEs rather was a (Carrollian) path-integral derivation of the CarrCFT EM tensor Ward identities and, from there, the Ward identities of the universal generators $S^+_0,S^+_1,T$ in \cite{Saha:2023hsl}. We speculate if this implies, in the absence of the operators that modify the subsubleading soft graviton theorem at the tree-level in the effective field theory \cite{Elvang:2016qvq}, that the semi-direct product of the (chiral) Virasoro algebra and the wedge sub-algebra of $\hat{w}_{1+\infty}$ is an exact quantum symmetry of the positive-helicity sector of any gravity-theory in $1+3$D bulk AFS, just like the specific case of the quantum self-dual gravity \cite{Ball:2021tmb}.      

\medskip 

In \cite{Banerjee:2023zip}, it was shown that there is a discrete infinite family of 2D Celestial CFTs possessing the (wedge sub-algebra of) $w_{1+\infty}$ symmetry. Two known examples of such theories are the MHV gravitons \cite{Banerjee:2020zlg,Banerjee:2021cly} and the quantum self-dual gravity \cite{Ball:2021tmb}. Since any $1+2$D CarrCFT containing the local field $S^+_2$ will enjoy the above said symmetry at the level of the OPEs, the conclusion of \cite{Banerjee:2023zip} suggests that there also exists an infinite number of $1+2$D CarrCFT of this type. It will be very interesting to construct such an explicit CarrCFT example.

\medskip

Finally, we showed that the CarrCFT Ward identity of the field $S^+_2$ with a special class of primaries \cite{Saha:2023hsl,Donnay:2022aba,Donnay:2022wvx,Nguyen:2023vfz} does indeed encode up to the subsubleading energetically soft graviton theorem \cite{Cachazo:2014fwa,Conde:2016rom}. Following this method, the infinite number of soft graviton theorems of \cite{Hamada:2018vrw,Li:2018gnc} were then directly interpreted as the Ward identities of the members of the infinite tower of Carrollian conformal primaries $S^+_k$. 

\medskip

In \cite{Saha:2023hsl}, it was found that the three universal CarrCFT generators $S^+_0,S^+_1,T$ inflict the ten global $\text{ISL}(2,\mathbb{C})$ Poincar\'e transformations on the quantum fields. Here, we showed that $S^+_2$ generates four additional global symmetry transformations that, unlike the Poincar\'e ones, are non-local in time (but local in space). The Einstein-gravity analogue of this result is described in \cite{Freidel:2021dfs}. We further clarified that the other primaries $S^+_{k\geq3}$ do not generate any further independent global symmetries. Recalling that the global symmetries constrain the correlators of a theory and the CarrCFT correlators can be mapped to the bulk AFS null-momentum space $S$-matrices \cite{Donnay:2022wvx}, our results provided a Carrollian justification of the statement \cite{Guevara:2021abz} that the infinite number of soft graviton theorems of \cite{Hamada:2018vrw,Li:2018gnc} beyond the subsubleading order does not impose any additional constraints on the bulk AFS mass-less $S$-matrices.

\medskip

An obvious future direction that can be pursued following the methodology presented in \cite{Saha:2023hsl} and in this work would be to figure out how the soft theorems of the gauge theories \cite{Hamada:2018vrw,Li:2018gnc,Campiglia:2018dyi} in the $1+3$D bulk AFS and the tower of conformally soft gluons of 2D Celestial CFT \cite{Guevara:2021abz} can arise in the framework of $1+2$D CarrCFT. By now it is apparent that some additional Carrollian conformal field(s) besides the three universal generators $S^+_0,S^+_1,T$ must be postulated to exist in the theory, the Ward identities of which would encode the soft gluon theorems. Similar situations have been considered in \cite{Zamolodchikov:1985wn} for 2D chiral CFTs and in \cite{Bagchi:2023dzx} for $1+1$D CarrCFTs.

\medskip

More important is to try to find a resolution to the problem of the double soft limits of opposite helicities \cite{Klose:2015xoa} within the CarrCFT framework. Extending the current work, one needs to start by assuming the existence of an opposite-spin counterpart of $S^+_2$. But as discussed in section \ref{s3}, this field $S^-_2$ can not be simultaneously treated as mutually local with $S^+_0,S^+_1,S^+_2,T$. The findings of the work \cite{Banerjee:2022wht} in the context of Celestial CFT are expected to play a very crucial role in this endeavour. We hope to report on this in a very near future. 

\medskip

\acknowledgments
I would like to thank Shamik Banerjee for illuminating discussions on Celestial holography at the beginning stage of this work, for a clarification on the work \cite{Banerjee:2023zip} and also for his valuable comments on the manuscript. It is a pleasure to acknowledge the warm hospitality of the National Institute of Science Education and Research (NISER), Jatni where this work was started. I am also grateful to Raju Mandal for many helpful discussions on Celestial CFT during my stay at NISER. I am greatly indebted to Alok Laddha and Romain Ruzziconi for emphasizing to me the correspondence between the large Carrollian time limit on the boundary and the energetically soft limit in the bulk. Finally, I thank Arjun Bagchi for his continued support. This work is financially supported by the PMRF fellowship, MHRD, India.

\end{document}